\documentclass[12pt]{article}

\usepackage{amsfonts,ulem}
\usepackage{amsmath}
\usepackage{amssymb,authblk}
\usepackage{amsthm}
\usepackage{array}
\usepackage{graphicx}
\usepackage[margin=1in]{geometry}
\usepackage{float}
\usepackage{booktabs}
 
\usepackage[config,labelfont=sc,textfont=sl]{caption,subfig}

\usepackage{mathtools}
\usepackage{psfrag}
\usepackage{mathrsfs}
 
\newcommand{\rmc}{\mathrm{c}} 
\newcommand{\rmd}{\mathrm{d}} 
\newcommand{\rme}{\mathrm{e}}

\newcommand{\rmi}{\mathrm{i}}

\newcommand{\rmm}{\mathrm{m}}

\newcommand{\rmp}{\mathrm{p}}

\newcommand{\rmB}{\mathrm{B}}

\newcommand{\rmH}{\mathrm{H}} 
 
\newcommand{\rmJ}{\mathrm{J}}

  \newcommand{\bfk}{\mathbf{k}}

\newcommand{\I}{\mathrm{i}}

\newcommand{\inc}{\mathrm{inc}}

 \usepackage{dutchcal}
\newcommand{\emm}{\mathbcal{m}} 
\newcommand{\qcal}{q} 

\title{\vspace{-20mm} Asymptotics of   the meta-atom: plane wave scattering by a single Helmholtz resonator }

\author[1]{M. J. A. Smith}
\author[2]{P. A. Cotterill} 
\author[3]{D. Nigro} 
\author[2]{W. J. Parnell} 
\author[1]{\\I. D. Abrahams} 

\affil[1]{\small Department of Applied Mathematics and Theoretical Physics, University of Cambridge,
Wilberforce Road, CB3 0WA, UK}
 \affil[2]{\small Department of Mathematics, The University of Manchester,  
  Oxford Road, Manchester, \newline M13 9PL, UK}
\affil[3]{\small Thales United Kingdom, 350 Longwater Avenue Green Park, Reading, RG2 6GF, UK}

\date{}





\begin{document}
\maketitle
  \vspace{-10mm}
\begin{center}
\small {\it Submitted Manuscript}
\end{center}	
\begin{abstract} \noindent
Using a combination of multipole methods and the method of matched asymptotics, we present a solution procedure for acoustic plane wave scattering by a single  Helmholtz resonator in two dimensions. Closed-form representations for the multipole scattering coefficients of the resonator are derived,  valid at low frequencies, with three fundamental configurations  examined in detail: the thin-walled, moderately thick-walled, and very thick-walled limits. Additionally, we examine the impact of   dissipation for   very thick-walled resonators, and also numerically evaluate the scattering, absorption, and extinction cross sections (efficiencies) for representative resonators in all three wall thickness regimes. In general, we observe   strong enhancement in both the scattered fields and cross sections at the Helmholtz resonance frequencies. As expected, dissipation is shown to  shift   the resonance frequency, reduce the amplitude of the field, and reduce the extinction efficiency at the fundamental Helmholtz resonance. Finally, we confirm   results in the literature on Willis-like coupling effects for this resonator design,  and crucially, connect these findings to earlier works by the authors on two-dimensional arrays of resonators, deducing that depolarisability effects (off-diagonal terms) for a single resonator do   not   ensure the existence of  Willis coupling effects (bianisotropy) in bulk.
\end{abstract}

\section{Introduction}
  {\it Wave scattering} is a process     describing how 
 an incident wave is modified by the presence of an object, or ensemble of objects, in its path \cite{bohren1998absorption,kerker1969scattering}. Factors such as the geometry of the scatterer or incident angle of the wave can induce major changes in the observed response, and using rigorous mathematical analysis, it is possible to provide insight on a diversity of observable effects and real-world phenomena. For example, wave scattering can be  used to identify  delaminations in   composite   media with  ultrasound   \cite{milton2002theory}, and explains how   water droplets in the air can give rise to vibrant rainbows  \cite{bohren1998absorption}. This broad  definition of scattering applies to waves   of all types, from surface   waves on deep water through to electromagnetic, elastic and sound waves,  and in general, there are   two canonical geometries whose scattering properties are now generally   well-understood: that of spheres and that of idealised (infinitely tall) circular cylinders.

There has been extensive research attention   on wave  scattering by spheres since at least the works of  Mie  \cite{mie1908beitage}, whose thorough outline  of the separation of variables solution for the vector Helmholtz equation continues to influence scientific research well into the present day \cite{bohren1998absorption,milton2002theory}. This is due to the fact that suspensions of spherical, or approximately spherical, objects arise in  an extensive range of applications,   from  the homogenisation of milk through to air pollution   modelling. Wave scattering by infinitely tall circular cylinders has likewise received  considerable attention, due to its extensive real-world applications, from the design of fibrous materials such as kevlar through to aerofoil design for aircraft \cite{milton2002theory}. It would appear that the first comprehensive multipole solution for scattering by circular cylinders, in the setting of Maxwell's equations, was conducted by von Ignatowsky \cite{v1905reflexion}, although the first  scattering solution was outlined  some decades earlier by Lord Rayleigh  \cite{rayleigh1881x}. An exhaustive early history of the scattering literature for spheres and cylinders may be found in Kerker \cite{kerker1969scattering}. Interest in multipole solutions for   cylinders has grown significantly in recent decades with the advent of two-dimensional photonic, phononic, and other crystals as well as with the advent of metamaterials (discussed below) \cite{joannopoulos2008molding,cui2010metamaterials}; however results are primarily obtained using numerical methods, with few analytical treatments available for non-circular cylindrical configurations.

In this work, we examine an important extension to scattering by an infinitely tall cylinder: we consider two-dimensional acoustic plane-wave  scattering by a circular cylinder  into which we have introduced  a hollow circular cavity and neck   to form a    Helmholtz resonator, where  Neumann boundary conditions are imposed on all walls. A representative resonator is presented in Figure \ref{fig:schematics} for guidance. Using multipole methods and the method of matched asymptotics \cite{crighton1992modern,cotterill2015time} we   construct a low-frequency, closed-form solution to describe the scattering coefficients for the resonator and the resulting potential field in the exterior domain. We are unaware of any other works in the literature which present low-frequency analytical expressions for the scattering coefficients of Helmholtz resonators in this manner. That said, scattering by a single thin-walled elastic Helmholtz resonator shell in a fluid has been considered previously \cite{norris1993elastic} and their resonance condition (Eq.~(C27)) in the rigid limit is equivalent to ours  in the  thin-walled setting. Here, we consider three resonator geometries explicitly: the thin-walled limit, moderately-thick walled limit, and extremely thick-walled limits, providing closed-form expressions and compact asymptotic forms in all instances.  

A key objective of the present work    is to  obtain   insight   on wave propagation through more complex arrangements of  resonators, for example, when they are tiled periodically to form finite clusters, gratings,  or lattices \cite{smith2022PartI,smith2022PartII}. The scatterer considered here is a canonical geometry for two-dimensional metamaterials, and following  an established nomenclature in the literature (i.e., Melnikov {\it et al.}  \cite{melnikov2019acoustic})   our resonator is  considered to be   a ``meta-atom''; this terminology arises from the fact that a two-dimensional periodic lattice of such resonators  can give rise to a {\it metamaterial}, a type of composite material that exhibits unusual and unexpected properties   that are not readily observed in conventional media \cite{joannopoulos2008molding,cui2010metamaterials}.

In addition to our multipole-matched asymptotic solution, we also compute     {\it cross sections} for  a single cylindrical Helmholtz resonator. These are measures of   strength for different wave processes, for example, describing how much incident power is scattered or absorbed by the resonator. We also evaluate extinction cross sections, which refer to the power loss in the downstream direction to the incident field (where a detector or observer would be located), due to the presence of the object \cite{bohren1998absorption}. Formally, extinction is defined as the sum of both absorption and scattering processes, and in particular frequency regimes, or for certain geometries, either scattering or absorption may dominate   \cite[Ch.~11]{bohren1998absorption}. In this work we consider   cross sections that are nondimensionalised by the geometric  cross section of a closed   cylinder and we term the resulting quantities {\it efficiency coefficients}. 
   Note that the extinction cross section may be used to determine an estimate for    the absorbance (attenuance)  of  a large random ensemble of resonators; this is   obtained by multiplying  the extinction cross section   by the filling fraction and the mean path length through the cluster (i.e., following the Beer-Lambert law \cite{swinehart1962beer}). A substantial body of literature   has been dedicated towards  cross-sections, particularly for spherical and cylindrical geometries, giving rise to significant variations in definitions  \cite{Baumeister1994scattering,bohren1998absorption,carney2004generalized,mechel2004formulas,norris2015acoustic,martin2018multiple}.  
In order to avoid any possible confusion, and to facilitate comparison with the literature, we present results for closed (ideal) Neumann cylinders where possible.

To further complement the above, we   examine the impact of dissipation by considering a boundary layer  in the neck region of a very thick-walled resonator.  A boundary layer refers to a wave-structure interaction effect where fluid viscosity effects begin to emerge near the surface of an object, ultimately dissipating wave energy. As discussed in the literature \cite{chester1981resonant,pierce2019acoustics}, there are a range of diffusive and dissipative mechanisms present in all  acoustic systems, such as bulk fluid thermal/viscous effects, radiation damping losses,   boundary layer effects, vortex shedding, and flow separation. In this work we consider     boundary layer effects (in the neck of   very thick-walled resonators only)  as we consider this to be the dominant internal loss mechanism.  We are able to show numerically that dissipation generally reduces the enhancement in the scattering and extinction cross sections (efficiencies) to recover   results for a closed Neumann cylinder, with the absorption efficiency exhibiting a clear peak as the attenuation coefficient $\bar{\alpha}$ increases. This absorption peak implicitly defines the operating range of our model, and we do not advise considering   $\bar{\alpha}$ values beyond this range, although   in the limit $\bar{\alpha}\rightarrow \infty$ results must return to those for a  closed Neumann cylinder, as the ever increasing resistance in the neck region inhibits energy flow to the interior of the resonator.

We remark that our closed form (multipole) solution representation  should prove useful to those interested in the scattering behaviour of   resonators, as our expressions are rapidly evaluable with little computational overhead. Our methodology avoids the need to use  intensive numerical methods (where meshing in the narrow neck region can become prohibitive), and permits the rapid exploration of parameter spaces for optimisation. To this end, we    present closed-form representations for the dispersion equation corresponding to the first Helmholtz resonance, in all three wall thickness regimes. In the very thick-walled regime,  the  fundamental Helmholtz resonance can be pushed to very low frequencies by tuning the wall thickness, outer radius, neck width, and neck length carefully (see Smith and Abrahams \cite{smith2022PartII}).

The outline of this paper is as follows. In Section \ref{sec:cylneumann} we present a brief solution outline for plane wave scattering by a closed (ideal) Neumann cylinder. Next,  we pose an ansatz for the total potential  of a  single  resonator satisfying Neumann conditions on the walls (which is known except for an undetermined coefficient) in Section \ref{sec:scatsinghelmres}. In Section \ref{sec:asymatch} we use matched asymptotic methods  to determine this amplitude for resonators in the  thin-walled, moderately thick-walled, and very thick-walled regimes. Closed-form expressions for determining the Helmholtz resonance frequency are then derived and presented for these three settings   in Section  \ref{sec:helmresfreq}. In Section \ref{eq:asyrepscatcoeff}, asymptotic representations for the scattering coefficients and definitions for the cross sections are   given, with dissipation in very thick-walled cylinders     considered in Section \ref{sec:diss}. Numerical results are   presented in Section \ref{sec:numerics} and are followed by concluding remarks in Section \ref{sec:disc}. Finally, in Appendix \ref{sec:willis} we discuss and clarify Willis-like coupling effects that have been reported in the literature for plane-wave scattering by   resonators of this or similar design.

\section{Scattering by a single cylinder}\label{sec:cylneumann}
As a means of reference, we begin by briefly outlining the multipole solution for time-harmonic  plane wave scattering by a single circular cylinder $\Omega_\rmc$ immersed in an infinitely extending fluid medium. We consider  the solution in the  domain exterior to the cylinder where the field satisfies the acoustic wave equation
\begin{equation}
\label{eq:helmholtzcyl}
( \partial_{ {x}}^2 + \partial_{ {y}}^2 + 1 ) \phi_\mathrm{ext} = 0,\quad\mbox{for} (x,y)\in \mathbb{R}_2 \backslash \Omega_\rmc,
\end{equation}
where  $x = k \bar{x}$ and $y=k\bar{y}$ denote nondimensional Cartesian coordinates, $\phi_\mathrm{ext}$ is the fluid velocity potential in the exterior domain, $k = \omega \sqrt{\rho/B}$ denotes the wavenumber,   $\omega$ is the angular frequency of the forcing and scattered fields, and $\rho$ and $B$ are the    density and bulk modulus    of the background medium, respectively. Here the observed time-dependent field is given by  $\mathrm{Re}\left\{\phi_\mathrm{ext}\exp(-\rmi \omega t) \right\}$, and the  general solution to \eqref{eq:helmholtzcyl} takes the form
\begin{equation}
\label{eq:phianbn}
\phi_\mathrm{ext} = \phi_\inc + \phi_\mathrm{sc} =  \sum_{n=-\infty}^{\infty} \left\{ a_n  \rmJ_n( r) +b_n  \rmH_n^{(1)}( r)\right\}\rme^{\rmi n \theta},
\end{equation}
where $\phi_\inc$ and $\phi_\mathrm{sc}$ are the incident and scattered potentials, respectively, $a_n $ and $b_n $  refer to the as yet unknown incoming and outgoing field coefficients, respectively, $\rmJ_n(z)$ is a Bessel functions of the first   kind, $\rmH_n^{(1)}(z)$ is a Hankel function of the first kind,  $(r,\theta)$ is the polar form of $(x,y)$, and we specify Neumann conditions on the walls
$
 \partial \phi_\mathrm{ext} /\partial r\big|_{r=b} = 0,
$
with  $\bar{b}$ denoting the dimensional   cylinder radius (and $b=k\bar{b}$).

 If we consider incident plane waves of the form $\phi_\inc = \exp( \rmi   x \cos \theta_\inc + \rmi  y\sin\theta_\inc )$, where $\theta_\inc$ denotes the incident angle,  then the solution    is obtained straightforwardly  by using a Jacobi--Anger expansion for the incident field \cite[Eq. (8.511-4)]{gradshteyn2014table} 
\begin{equation}
\label{eq:jacobianger}
\rme^{\rmi r \cos\theta} = \sum_{n=-\infty}^{\infty} \rmi^n \rmJ_n(r) \rme^{\rmi n \theta},
\end{equation}
  and   imposing the Neumann boundary condition, which gives the form
\begin{equation}
\label{eq:gensolcylinders}
\phi_\mathrm{ext} =
 \rme^{ \rmi   r\cos(\theta-\theta_\inc )}
 -
\sum_{n=-\infty}^{\infty}  \left\{    \frac{ \rmi^n \rmJ_n^\prime( b)}{\rmH_n^{(1)\prime}(b)} \rme^{-\rmi n \theta_\inc} \right\} \rmH_n^{(1)}( r) \, \rme^{\rmi n \theta },
\end{equation}
where prime notation denotes the derivative with respect to argument, i.e., $\rmJ_n^\prime(b) = \partial_z \rmJ_n(z)|_{z=b}$.
With this canonical   solution for plane wave scattering by a cylinder  in mind, we now proceed to our solution outline for a  cylindrical resonator. 

\section{Scattering by a single Helmholtz resonator}\label{sec:scatsinghelmres}
In order to construct a multipole solution for a single resonator at low frequencies, we  combine  the multipole expansion techniques outlined in Section \ref{sec:cylneumann} above,  with the method of matched asymptotic expansions \cite{crighton1992modern,cotterill2015time}. This procedure  is outlined in extensive detail for two-dimensional homogeneous {\it arrays} of resonators in earlier works by the authors  \cite{smith2022PartI,smith2022PartII}; as the inner solutions and matching procedure are unchanged from these earlier works, we outline only  the updated (leading-order) outer solutions  here, and simply state key results   where appropriate. For reference, the acoustic wave equation \eqref{eq:helmholtzcyl} governs wave propagation in the exterior domain for all outer solutions.

\subsection{Leading-order outer solution for all wall thicknesses} \label{sec:loos}
\begin{figure}[t]
\centering
\subfloat[Subfigure 6 list of figures text][]{
\includegraphics[width=0.495\textwidth]{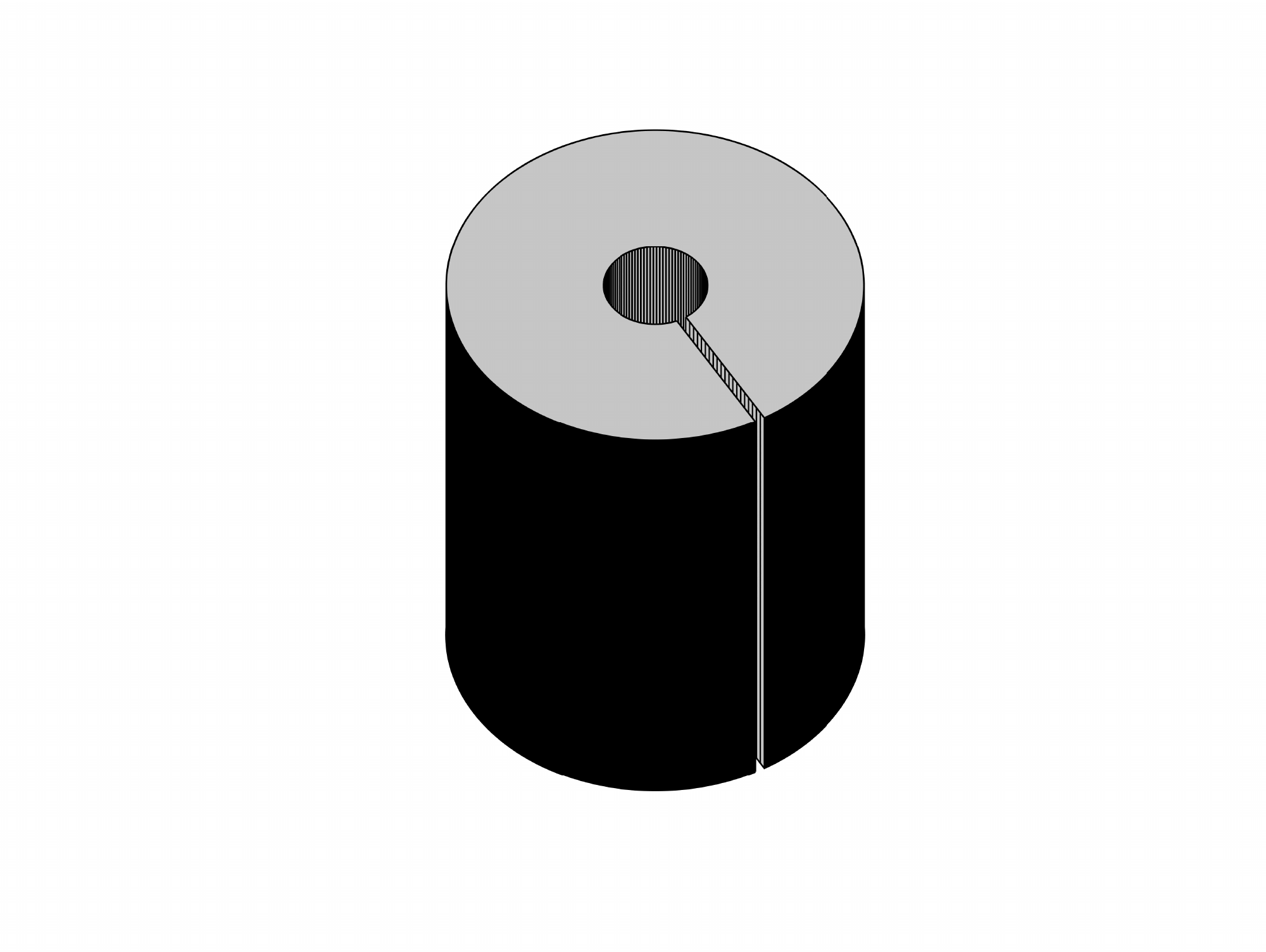}
\label{fig:figxx1}}
\subfloat[Subfigure 1 list of figures text][]{
\includegraphics[width=0.405\textwidth]{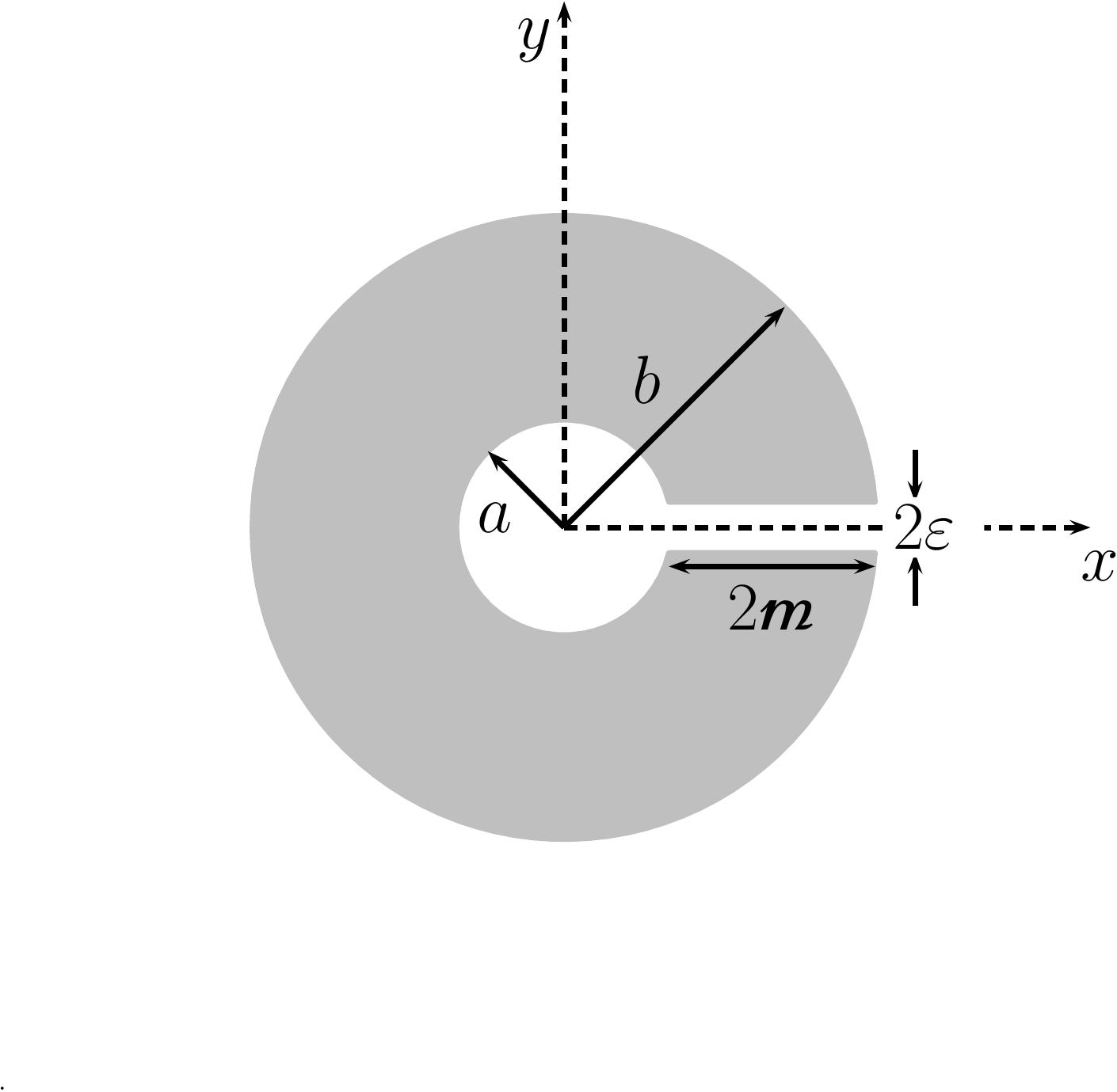}
\label{fig:figxx2}}

\caption{  \protect\subref{fig:figxx1}  Three-dimensional segment  and \protect\subref{fig:figxx2} two-dimensional cross section of our infinitely tall Helmholtz resonator in dimensionless coordinates. }
\label{fig:schematics}
\end{figure}

We begin by posing an ansatz for the   exterior domain ($r \geq b$), comprising the plane wave,  a monopole source   at the (small) resonator mouth, and a complete cylindrical harmonic basis,   in the form
\begin{equation}
\label{eq:ansatz1}
\phi_\mathrm{ext} = \rme^{\rmi   r\cos(\theta-\theta_\inc )}+A \rmH_0^{(1)}( \tilde{r}) + \sum_{n=-\infty}^{\infty}  c_n      \rmH_n^{(1)}( r) \rme^{\rmi n \theta},
\end{equation}
where $A$ and $c_n $ are  as yet unknown coefficients,  $\tilde{r} = \sqrt{(x-x_0)^2+(y-y_0)^2}$, and $(x_0,y_0) = (b \cos\theta_0,b\sin\theta_0)$ represents the central location of the resonator mouth. Without loss of generality we now fix the location of the aperture by taking $\theta_0 = 0$ and consider only varying incident angle $\theta_\inc$ in the present work. Figure \ref{fig:schematics} indicates the geometry of the scatterer.

For our leading-order outer problem \cite{crighton1992modern,smith2022PartI,smith2022PartII}, the resonator is an {\it almost-closed} cylinder, and so we treat the aperture as a point source via the   boundary condition     \cite{smith2022PartI}
\begin{equation}
\label{eq:dirich1}
\frac{\partial \phi_\mathrm{ext}}{\partial r}\bigg|_{r=b} = \frac{\rmi A}{\pi b} \sum_{n=-\infty}^{\infty} \rme^{\rmi n  \theta  },
\end{equation}
which is compatible with the   point source term in \eqref{eq:ansatz1}. 
After applying Graf's addition theorem \cite[Eq. (8.530)]{gradshteyn2014table} 
\begin{equation}
\label{eq:gadh0}
   \rmH_0^{(1)}(\widetilde{r})
   =
   \sum\limits_{n=-\infty}^{\infty} \rmJ_n(b) \rmH_n^{(1)}(r)\rme^{\rmi n (\theta - \theta_0)}, \quad \mbox{for} \quad r>b, 
\end{equation}
on the initial monopole term in \eqref{eq:ansatz1}, and the Jacobi--Anger expansion   \eqref{eq:jacobianger} on the plane wave term, we impose the Neumann condition  \eqref{eq:dirich1} above to obtain  
\begin{equation}
\label{eq:ancoeff}
c_n  =-\rmi^n \frac{\rmJ^\prime_n(b)}{\rmH_n^{(1)\prime}(b)} \rme^{-\rmi n \theta_\inc} - \frac{A Q_n}{2\rmH_n^{(1)\prime}(b)},
\end{equation}
where   $Q_n = \rmJ_n(b) \rmH_n^{(1)\prime} (b)+\rmJ_n^\prime(b) \rmH_n^{(1)} (b) $. Accordingly, only the  amplitude $A$ remains unknown at this stage, and in the limit as we approach the exterior mouth we find that 
  \begin{multline}
  \label{eq:phiextnearap}
\lim_{r\rightarrow b}\lim_{\theta\rightarrow  0}\phi_\mathrm{ext}\sim
\rme^{\rmi b \cos \theta_\inc} 
+ \frac{2 \rmi A }{\pi} \left( \gamma_\rme - \frac{\rmi \pi}{2} 
+ \log\left( \frac{\tilde{r}}{2}\right)\right) 
\\
 - \sum_{n=-\infty}^{\infty} \left[ \rmi^n \rmJ_n^\prime(b) \rme^{-\rmi n \theta_\inc} + \frac{A Q_n }{2}\right]\frac{ \rmH_n^{(1)}(b)}{\rmH_n^{(1)\prime}(b)},
\end{multline}
where $\gamma_\rme \approx 0.577216\ldots$ denotes the Euler--Mascheroni constant.  Having derived an appropriate asymptotic form for the outer solution   in the exterior domain \eqref{eq:phiextnearap}, as we approach the aperture, we   now consider asymptotic matching   to determine $A$ at low frequencies. 
 
\subsection{Results from asymptotic matching } \label{sec:asymatch}
By employing the   asymptotic matching procedure   outlined  in   \cite{smith2022PartI,smith2022PartII}, where expressions for all other leading-order outer solutions (i.e., for the resonator interior $\phi_\mathrm{int}$ and the resonator neck $\phi_\mathrm{neck}$) and all inner solutions (i.e., $\Phi$ and $\Psi$) are given explicitly, we now use the   representation for $\phi_\mathrm{ext}$ in \eqref{eq:phiextnearap} of the present work, and find   after  considerable algebraic manipulation that the amplitude is given by
\begin{equation}
\label{eq:crucA}
A = -\frac{2\rmi}{\pi    b  h_\varepsilon} \sum_{p=-\infty}^{\infty} \frac{\rmi^p }{\rmH_p^{(1)\prime}( b)}\rme^{-\rmi p \theta_\inc},
\end{equation}
where  
\begin{equation}
\label{eq:hepsall}
 h_\varepsilon = 
\begin{cases}
\dfrac{4\rmi}{\pi}\left( \gamma_\rme - \dfrac{\rmi \pi}{2} + \log\left(\dfrac{\varepsilon}{4}\right)\right) - \dfrac{1}{2}\sum\limits_{m=-\infty}^{\infty} \dfrac{Q_m^2}{\rmJ_m^\prime(b) \rmH_{m}^{(1)\prime}(b)}, \quad \mbox{when thin-walled,  } & \\[12pt]
\dfrac{4\rmi}{\pi}\left( \gamma_\rme - \dfrac{\rmi \pi}{2} + \log\left(\dfrac{\sqrt{\qcal}\, \mathcal{C}(\qcal)\varepsilon}{2}\right)\right) - \dfrac{1}{2} \sum\limits_{m=-\infty}^{\infty} \left\{ \dfrac{\check{Q}_m \rmJ_m(a)}{\rmJ_m^\prime(a)}+\dfrac{Q_m \rmH_m^{(1)}(b)}{\rmH_m^{(1)\prime}(b)} \right\}, \qquad & \\
\hspace{65mm}\mbox{when moderately thick-walled}, &\\[12pt]
\dfrac{2\rmi}{\pi}\left( \gamma_\rme - \dfrac{\rmi \pi}{2} - \log\left(\dfrac{\pi}{\varepsilon}\right) - \left[ \dfrac{2\rmi \tau_3}{\pi} + \tau_4\tau_5\right] \left[  \dfrac{2\rmi \tau_1}{\pi} + \tau_2\tau_5\right]^{-1}	\right) - \dfrac{1}{2}\sum\limits_{m=-\infty}^{\infty}	 \dfrac{Q_m \rmH_m^{(1)}(b)}{\rmH_m^{(1)\prime}(b)} , \qquad & \\[12pt]
\hspace{75mm} \mbox{when very thick-walled.  }
\end{cases} 
\end{equation} 
In these expressions, we introduce the small dimensionless parameter $\varepsilon = k\bar{\ell}$, where $2\bar{\ell}$  denotes the total aperture width, $2\bar{\emm}$ is the total aperture length (where $\emm = k\bar{\emm}$), $a = b- 2\emm$ denotes the inner radius (where $a = k\bar{a}$),  $\check{Q}_m = \rmJ_m(a) \rmH_m^{(1)\prime} (a)+\rmJ_m^\prime(a) \rmH_m^{(1)} (a) $, and the quantity $\qcal$ is obtained by solving the relation \cite{smith2022PartI}
  \begin{equation}
\label{eq:scmapqh}
\frac{\emm}{\ell} = \frac{1}{2}\left[ 2 E(\qcal^2) + (\qcal^2-1) K(\qcal^2)\right]^{-1}\left[- 2E(1-\qcal^2) + (1+\qcal^2) K(1-\qcal^2)  \right],
\end{equation}
with
$
\mathcal{C}(\qcal) =1/ \{ 2 E(\qcal^2) + (\qcal^2-1)K(\qcal^2)  \},
$
where $E(z)$ and $K(z)$ are complete Elliptic integrals of the first and second kind, respectively.   Furthermore, as in Smith and Abrahams \cite{smith2022PartII} we define  
\begin{subequations}
\label{eq:listotaus}
\begin{align}
\tau_1 &=  \frac{2\varepsilon}{\pi} (1-\log 2) \sin(2\emm) - \cos(2\emm), \qquad 
\tau_4  =  -\frac{2\varepsilon}{\pi} (1 - \log 2) \sin(2\emm)+ \cos(2\emm), 
\\
\tau_2 &= -\frac{2\varepsilon}{\pi}\sin(2\emm), \hspace{20mm}
\tau_5  = \dfrac{2\rmi  }{\pi} \left[ \gamma_\rme - \dfrac{\rmi \pi}{2}  -\log\left(\frac{\pi}{ \varepsilon}\right) \right] - \dfrac{1}{2}\sum\limits_{m=-\infty}^{\infty} \dfrac{\check{Q}_m \rmJ_m(a)}{\rmJ_m^\prime(a)}   ,
\\
\tau_3 &=  \left[ \frac{2\varepsilon}{\pi} (1-\log 2)^2   - \frac{\pi}{2\varepsilon}\right]\sin(2\emm) - 2 (1-\log 2)  \cos(2\emm).  
\end{align}
\end{subequations}
Thus, with the form of $A$ in \eqref{eq:crucA} and the   representation for $h_\varepsilon$ in \eqref{eq:hepsall}, in addition to the expression for $c_n $ in \eqref{eq:ancoeff}, we    have  now fully prescribed      $\phi_\mathrm{ext}$ in \eqref{eq:ansatz1} in the low-frequency asymptotic limit for all wall thickness configurations. For reference, we also define the channel aspect ratio $h = 2\bar{\emm}/2\bar{\ell}$ and note that  in the thin-walled limit $h   \rightarrow 0$, we have $\qcal\rightarrow 1$ and $\mathcal{C}(\qcal) \rightarrow 1/2$, which returns   consistent expressions for $h_\varepsilon$ in \eqref{eq:hepsall}. 

Next, we   examine the explicit form of the Helmholtz resonance condition for a single resonator.

\subsection{Helmholtz resonance frequency}\label{sec:helmresfreq}
For our plane  wave scattering problem, the Helmholtz resonance frequency is obtained straightforwardly by searching  for the minimum value of the denominator of $A$ in \eqref{eq:crucA}. This corresponds to the vanishing of the imaginary part of $h_\varepsilon$. In the limit as the wave becomes long relative to all geometric parameters,  we find that
\begin{equation}
\label{eq:allbigheps}
 \lim_{ \left\{a,b,\emm \right\}\rightarrow 0}   h_\varepsilon \approx 
\begin{cases}
\dfrac{ \rmi}{\pi}\left[  \dfrac{2}{b^2} - \dfrac{1}{4} + \gamma_\rme - \dfrac{\rmi \pi}{2} + \log\left(\dfrac{\varepsilon^4}{2^5 b^3}\right)  \right], \quad \mbox{when thin-walled,  } & \\[12pt]
\dfrac{ \rmi}{\pi}\left[  \dfrac{2}{a^2} - \dfrac{1}{4} + \gamma_\rme - \dfrac{\rmi \pi}{2} + \log\left(\dfrac{\qcal^2 \mathcal{C}(\qcal)^4\varepsilon^4}{2 a^2 b}\right)  \right], \quad   \mbox{when moderately thick-walled}, &\\[12pt]
\dfrac{ \rmi}{\pi} \left\{ \gamma_\rme  - \dfrac{\rmi\pi}{2} -  \log\left(\dfrac{\pi^2 b}{2\varepsilon^2}\right) + 2 \left[ \dfrac{1}{a^2} - \dfrac{\emm\pi}{\varepsilon} - \dfrac{17}{8} - \log\left(\dfrac{\pi a}{2^3\varepsilon}\right)\right] \right.	   \qquad & \\[12pt]
\hspace{10mm} \cdot\left.\left[1 + \dfrac{4\varepsilon \emm}{\pi}\left( \dfrac{1}{a^2} - \dfrac{1}{8} - \log\left(\dfrac{\pi a}{2\varepsilon}\right)\right)\right]^{-1}\right\}, \quad \mbox{when very thick-walled.  }
\end{cases} 
\end{equation} 
If we then prescribe $\emm =\kappa_\emm \varepsilon^\mu$ and $a = \kappa_a \varepsilon^\gamma$ where $\kappa_\emm$ and $\kappa_a$ are real constants and $0<\mu,\gamma<1$, and   examine the narrow aperture limit $\varepsilon\rightarrow 0$, it follows that under the dominant balance scaling $\mu + 1 - 2\gamma>0$ \cite{smith2022PartII} we have
\begin{equation}
h_\varepsilon \approx \dfrac{ \rmi}{\pi} \left\{  \dfrac{2}{a^2}   - \dfrac{2\emm\pi}{\varepsilon}   - \dfrac{17}{4}  + \gamma_\rme - \dfrac{\rmi\pi}{2}     + \log\left(\dfrac{2^7\varepsilon^4}{\pi^4  a^2 b}\right)     \right\}	  , \quad \mbox{when very thick-walled. }
\end{equation}
Accordingly we write the Helmholtz resonance condition in our three canonical limits  as
\begin{subequations}
\label{eq:allhelmresexpr}
\begin{align}
\label{eq:helmresthin}
    \dfrac{2}{b^2} - \dfrac{1}{4} + \gamma_\rme   + \log\left(\dfrac{\varepsilon^4}{2^5 b^3}\right)  &\approx 0, \quad \mbox{when thin-walled,  }  \\ 
    \label{eq:helmresmodthick}
   \dfrac{2}{a^2} - \dfrac{1}{4} + \gamma_\rme  + \log\left(\dfrac{\qcal^2 \mathcal{C}(\qcal)^4\varepsilon^4}{2 a^2 b}\right) &\approx  0, \quad   \mbox{when moderately thick-walled}, \\ 
   \label{eq:helmdispeqvthick}
   \dfrac{2}{a^2}   - \dfrac{2\emm\pi}{\varepsilon}  - \dfrac{17}{4}  + \gamma_\rme      + \log\left(\dfrac{2^7\varepsilon^4}{\pi^4  a^2 b}\right)   &\approx  0	  , \quad \mbox{when very thick-walled, }
\end{align}
\end{subequations}
and note that in all cases $h_\varepsilon$ takes the value $1/2$ at resonance. These expressions must be solved numerically due to the presence of $\log(k)$ terms, in contrast to the corresponding expressions for a two-dimensional array of resonators \cite{smith2022PartI,smith2022PartII}.

The   expressions in \eqref{eq:allhelmresexpr} are useful for rapidly determining the location of the first Helmholtz resonance. For example, should we specify a total  aperture width of 	 $2\bar{\ell} \approx 1\, $mm, which generally corresponds to an acoustic length scale  before viscous effects   start to emerge, as well as a desired resonance  frequency $\nu$   (i.e., $\nu = 1 $ kHz, where $\omega = 2\pi \nu$ or  $k = 18.3074$ m$^{-1}$ in air) then in the thin-walled limit we solve  \eqref{eq:helmresthin} to find that an outer radius of $\bar{b} \approx 17.8$ mm is required; such values correspond to a resonator design that is readily fabricated using contemporary methods (i.e., using additive manufacturing techniques).

\section{Asymptotic representations of scattering coefficients} \label{eq:asyrepscatcoeff}
In this section, we return to the leading-order exterior field representation from Section \ref{sec:loos} and examine the multipole coefficients in closer detail. We wish to obtain simple explicit expressions valid at low frequencies, i.e., for $b=k\bar{b}\ll 1$, ensuring that we preserve the partitioning $\varepsilon\ll b$. To this end, we write the  ansatz \eqref{eq:ansatz1}  as
\begin{equation}
\label{eq:expanphiext2}
\phi_\mathrm{ext} = \phi_\inc + \phi_\mathrm{sc} =  \rme^{\rmi   r\cos(\theta-\theta_\inc )}  + \sum_{n=-\infty}^{\infty}  d_n      \rmH_n^{(1)}( r) \rme^{\rmi n \theta},
\end{equation}
which arises after an application of Graf's addition theorem \eqref{eq:gadh0}, where
\begin{equation}
\label{eq:bnfinale}
d_n  = A \rmJ_n(b) + c_n  =  \frac{\rmi A}{\pi b \rmH_n^{(1)\prime}(b)}   - \rmi^n \frac{\rmJ^\prime_n(b)}{\rmH_n^{(1)\prime}(b)} \rme^{-\rmi n \theta_\inc} .
\end{equation}
On inspection,  we see that the first term in \eqref{eq:bnfinale} is due to  the   resonator   and   the second term corresponds to the scattering coefficient for an ideal Neumann cylinder  \eqref{eq:gensolcylinders}.
Next, we   truncate the sum in \eqref{eq:expanphiext2} so that all terms of $O(b^4)$ are captured (i.e., we consider the orders $n = -2,\ldots,2$) and  presume that  such a truncation is sufficient   for describing the   response of the resonator at low frequencies. Consequently,   in the vanishing $b$ limit   we find that
\begin{subequations}
\label{eq:asyformsAqns}
\begin{align}
\label{eq:helmreseq}
\lim_{b \rightarrow 0}
d_0  &\approx  \frac{\pi \rmi b^2}{8f_\varepsilon}\left[1 -  2 f_\varepsilon \vphantom{\frac{1}{2}} \right] - \frac{\pi b^3\cos(\theta_\inc) }{4 f_\varepsilon} \\
&\hspace{10mm}+ \frac{\pi \rmi b^4}{8}\left\{ \left(1 - \frac{1}{f_\varepsilon} \right) \left[ \frac{1}{2} - \gamma_\rme + \frac{\rmi \pi}{2} - \log\left(\frac{b}{2}\right) \right] + \frac{1}{4} - \frac{\cos(2\theta_\inc)}{2 f_\varepsilon}  \right\} ,    \\
\lim_{b \rightarrow 0}
d_{\pm 1}  &\approx  \mp \frac{\pi b^2 }{4}\rme^{\mp\rmi\theta_\inc} \pm \frac{\pi \rmi b^3}{8f_\varepsilon}   + \frac{\pi b^4 }{4} \left\{\pm \frac{\rme^{\mp \rmi\theta_\inc}}{2} \left[ \frac{5}{4} + \gamma_\rme - \frac{\rmi \pi}{2} + \log\left(\frac{b}{2}\right)\right] \mp \frac{\cos(\theta_\inc)}{f_\varepsilon} \right\}, \\ \nonumber
\lim_{b \rightarrow 0} d_{\pm 2}  &\approx  \frac{\pi \rmi b^4}{32}\left[ \frac{1}{f_\varepsilon} - \rme^{\mp 2\rmi\theta_\inc}\right] , 
\end{align}
\end{subequations}
where  we have used the scaling $h_\varepsilon = 4\rmi f_\varepsilon/(\pi b^2)$ and the result
\begin{equation}
\lim_{b \rightarrow 0}A
  \approx  \frac{\pi \rmi b^2}{4 f_\varepsilon} \left(1 + 2 \rmi b \cos(\theta_\inc)\right)
  + \frac{\rmi \pi b^4}{8f_\varepsilon}\left[  -\frac{1}{2} + \gamma_\rme - \frac{\rmi \pi}{2} - \cos(2\theta_\inc) + \log\left(\frac{b}{2}\right)\right].
\end{equation}
 With these representations in mind, we now consider   cross sections for our Helmholtz resonator.

\subsection{Scattering, absorption, and extinction cross sections}\label{sec:xsections}
In order to compactly describe how our resonator influences the incident plane wave, we evaluate cross sections for the scattering, absorption and extinction strength of a given resonator, which are all functions of frequency. We characterise the     scattering strength using  the (dimensional)  scattering cross section  for a   cylinder or resonator     \cite[Sec.~E.1]{mechel2004formulas}
\begin{equation}
\label{eq:sigmabarscatdef}
 \bar{\sigma}_\mathrm{sc}  = \frac{4}{ k }\sum_{m=-\infty}^{\infty} |d_m |^2,
\end{equation}
which we scale by the geometric cross section  of the scatterer to define the nondimensional    coefficient
$
Q_\mathrm{sc} =  \bar{\sigma}_\mathrm{sc}/(2 \bar{b}),
$
which is termed the {\it scattering efficiency}  \cite{bohren1998absorption}. Additionally we have the  (dimensional)  extinction cross section   
\begin{equation}
\label{eq:sigmaextbar}
 \bar{\sigma}_\mathrm{ext}  = -\frac{4}{k}\,   \sum_{m=-\infty}^{\infty} \mathrm{Re} \left\{     d_m  \rme^{-\rmi m (\pi/2 -  \theta_\inc) } \right\},    
\end{equation}
with the corresponding extinction efficiency $
Q_\mathrm{ext} =  \bar{\sigma}_\mathrm{ext}/(2\bar{b})
$. Furthermore, we note that the extinction efficiency is given by the sum \cite{bohren1998absorption}:
\begin{equation}
\label{eq:sumQs}
Q_\mathrm{ext}  = Q_\mathrm{sc} + Q_\mathrm{abs},
\end{equation}
where $Q_\mathrm{abs} = \bar{\sigma}_\mathrm{abs}/(2\bar{b})$ denotes the   absorption efficiency. Therefore, in the absence of viscosity (or other dissipative processes) we have that $Q_\mathrm{ext} =Q_\mathrm{sc}$. Having obtained cross section expressions    in preparation for    numerical investigations in Section \ref{sec:numerics}, we now examine the impact of dissipation in   thick-walled resonators.

\section{Dissipation in very thick-walled resonators}\label{sec:diss}
In this section, we   discuss the impact of   incorporating viscosity within the neck region of a very thick-walled resonator, where     a boundary layer could be expected to appear in the fluid, giving rise to viscous losses   \cite{pierce2019acoustics}. For thin- and moderately thick-walled resonators, boundary layer effects can be expected to be minimal, due to the smaller neck length, and so we do not consider   these configurations here. To incorporate dissipative loss in a simple fashion, we   specify  a complex-valued wavenumber  within the   neck region via the replacement $k \mapsto k + \rmi \bar{\alpha}$ where $\bar{\alpha}$ is the dimensional attenuation coefficient (units $1/\rmm$). This  gives rise to the outer solution in the neck (cf., \cite[Eq.~(2.6)]{smith2022PartII})
\begin{equation}
\label{eq:phineck}
\lim_{\varepsilon\rightarrow 0}\phi_\mathrm{neck} \sim p_0 \rme^{\rmi (1+\rmi \alpha)  \tilde{y}} +q_0 \rme^{-\rmi  (1+\rmi \alpha) \tilde{y}},
\end{equation}
where $\alpha = \bar{\alpha}/k$ is the nondimensional attenuation constant,    $p_0$ and $q_0$ are unknown constants, and $(\tilde{x},\tilde{y})$ denote a local rotated coordinate frame whose origin is centered at the exterior mouth of the resonator \cite{smith2022PartII}. Using the   outer solution  in the exterior domain $\phi_\mathrm{ext}$   \eqref{eq:phiextnearap}, the outer solution in the neck region $\phi_\mathrm{neck}$ \eqref{eq:phineck}, and the outer solution in the interior \cite[Eq.~(6.11)]{smith2022PartI}, in tandem with the inner solutions   $\Phi$ and $\Psi$  given in Smith and Abrahams \cite{smith2022PartII}, we obtain   a result identical to \eqref{eq:crucA}  after asymptotic matching, but with the replacement $h_\varepsilon \mapsto h_\varepsilon^\rmd$  where
\begin{equation}
\label{eq:hepsdislossy}
h_\varepsilon^\rmd = \dfrac{2\rmi}{\pi}\left( \gamma_\rme - \dfrac{\rmi \pi}{2} - \log\left(\dfrac{\pi}{\varepsilon}\right) - \left[ \dfrac{2\rmi \tau_3^\rmd}{\pi} + \tau_4^\rmd\tau_5\right] \left[  \dfrac{2\rmi \tau_1^\rmd}{\pi} + \tau_2^\rmd\tau_5\right]^{-1}	\right) - \dfrac{1}{2}\sum\limits_{m=-\infty}^{\infty} \dfrac{Q_m \rmH_m^{(1)}(b)}{\rmH_m^{(1)\prime}(b)} 	, 
\end{equation}
 along with  the dissipative forms
\begin{subequations}
\label{eq:lossytausall}
\begin{align}
\tau_1^\rmd &=  \frac{2\varepsilon \gamma_\rmd}{\pi} (1-\log 2) \sin(2\emm\gamma_\rmd) - \cos(2\emm\gamma_\rmd), 
\\ 
\tau_2^\rmd &= -\frac{2\varepsilon\gamma_\rmd}{\pi}\sin(2\emm\gamma_\rmd), 
\\
\tau_3^\rmd  &=  \left[ \frac{2\varepsilon\gamma_\rmd}{\pi} (1-\log 2)^2   - \frac{\pi}{2\varepsilon\gamma_\rmd}\right]\sin(2\emm\gamma_\rmd) - 2 (1-\log 2)  \cos(2\emm\gamma_\rmd), 
\\
\tau_4^\rmd  &=  -\frac{2\varepsilon\gamma_\rmd}{\pi} (1 - \log 2) \sin(2\emm\gamma_\rmd)+ \cos(2\emm\gamma_\rmd), 
\end{align}
\end{subequations}
in which $\gamma_\rmd = 1 + \rmi \alpha$.  Presuming that $\alpha$ is small (i.e., $\alpha \sim O(1)$) we have that
\begin{multline}
\label{eq:lossyhepsasy}
 \lim_{ \left\{a,b,\emm \right\}\rightarrow 0}   h_\varepsilon^\rmd \approx  \frac{2\rmi}{\pi} \left\{ \frac{\gamma_\rme}{2} - \frac{\rmi \pi}{4} - \frac{1}{2}\log\left(\frac{\pi^2 b}{2 \varepsilon^2}\right) \right. \\ \left.
 + \left[ \frac{1}{a^2} - \frac{\emm \pi}{\varepsilon} - \frac{2\pi\emm^3\gamma_\rmd^2}{3\varepsilon} - \frac{2 \emm^2 \gamma_\rmd^2}{a^2} - \frac{17}{8} - \log\left(\frac{\pi a}{8\varepsilon} \right)\right]  \cdot \left[ 1 - 2\emm^2\gamma_\rmd^2 + \frac{4 \varepsilon \emm \gamma_\rmd^2}{\pi a^2}\right]^{-1} \right\},
 \end{multline}
which, under  the $\mu+1-2\gamma>0$ dominant balance limit from Section  \ref{sec:helmresfreq},  takes the form
\begin{equation}
\label{eq:dombalhvisc}
h_\varepsilon^\rmd  \approx \frac{\rmi}{\pi}\left\{ \gamma_\rme - \frac{\rmi \pi}{2} - \frac{17}{4} + \frac{2}{a^2} - \frac{2\emm \pi}{\varepsilon} + \log \left( \frac{2^7 \varepsilon^4}{\pi^4 a^2 b}\right) - 4 \emm^2 \gamma_\rmd^2 \left[  \frac{ 1 }{a^2} +  \frac{  \emm \pi}{3\varepsilon}   \right]  \right\},
\end{equation}
and subsequently, the Helmholtz resonance condition is given by
\begin{equation}
\label{eq:helmresvisc}
   \frac{2}{a^2} - \frac{2\emm \pi}{\varepsilon}   - \frac{17}{4}  + \gamma_\rme  + \log \left( \frac{2^7 \varepsilon^4}{\pi^4 a^2 b}\right) - 4 \emm^2 (1  - \alpha^2) \left[  \frac{ 1 }{a^2} +  \frac{  \emm \pi}{3\varepsilon}   \right]  =0.
\end{equation}
Accordingly, by substituting \eqref{eq:helmresvisc} into \eqref{eq:dombalhvisc} we find that {\it at} the Helmholtz resonance we have
\begin{equation}
\label{eq:hepsdonres}
h_\varepsilon^\rmd \approx
     \frac{  1}{2} +  \frac{8}{\pi} \alpha  \emm^2   \left[  \frac{ 1 }{a^2} +  \frac{  \emm \pi}{3\varepsilon}   \right]  ,
\end{equation}
and so from the definition of $A$ in \eqref{eq:crucA} (with the replacement $h_\varepsilon \mapsto h_\varepsilon^\rmd$) we see that the presence of dissipation acts to lower the amplitude of the monopole source when on-resonance. That is, $h_\varepsilon^\rmd$ increases from the value of $1/2$ observed in the lossless case to \eqref{eq:hepsdonres}. 
 For reference, we use  $ h_\varepsilon^\rmd =  4\rmi  f_\varepsilon^\rmd / ( \pi b^2)$  for the replacement  $f_\varepsilon \mapsto f_\varepsilon^\rmd$. Furthermore, the above forms   for $h_\varepsilon^\rmd$ \eqref{eq:hepsdislossy} and  $\tau_j^\rmd$ \eqref{eq:lossytausall} correctly limit to the earlier nondissipative results for $h_\varepsilon$ \eqref{eq:hepsall} and $\tau_j$ \eqref{eq:listotaus} in the limit of  vanishing  loss  $\alpha \rightarrow 0$.

\section{Numerical Results} \label{sec:numerics}
In this section, we compute a selection of scattered field profiles $\phi_\mathrm{sc}$ for representative resonators from all three wall thickness configurations,  numerically evaluate  a selection of scattering, absorption, and extinction efficiencies  as a function of  nondimensionalised  frequency (wavenumber), and examine how well the   asymptotic representations for $d_n $ in \eqref{eq:asyformsAqns} perform relative to the full multipole forms \eqref{eq:bnfinale}. Additionally, we consider the impact of  dissipative loss (from a boundary layer in the neck)   on all cross sections for a representative resonator in the  very thick-walled limit. In all examples we consider   air as the background medium possessing a bulk modulus $B=141.83$ KPa and density $\rho = 1.2041$ kg/m$^3$, although  this is done without loss of generality.

\begin{figure}[t]
\centering
\subfloat[Subfigure 6 list of figures text][]{
\includegraphics[width=0.495\textwidth]{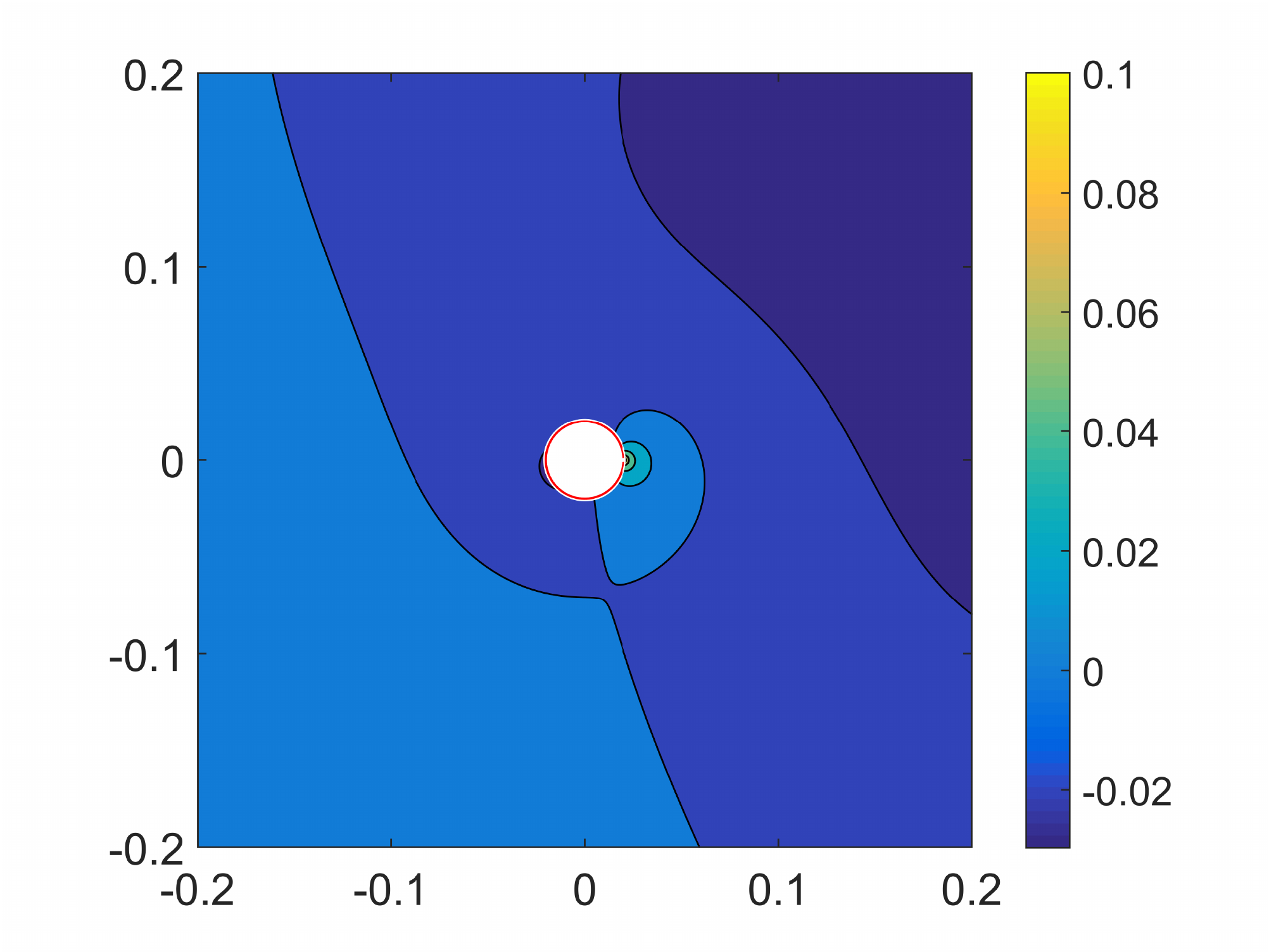}
\label{fig:figa1}}
\subfloat[Subfigure 1 list of figures text][]{
\includegraphics[width=0.495\textwidth]{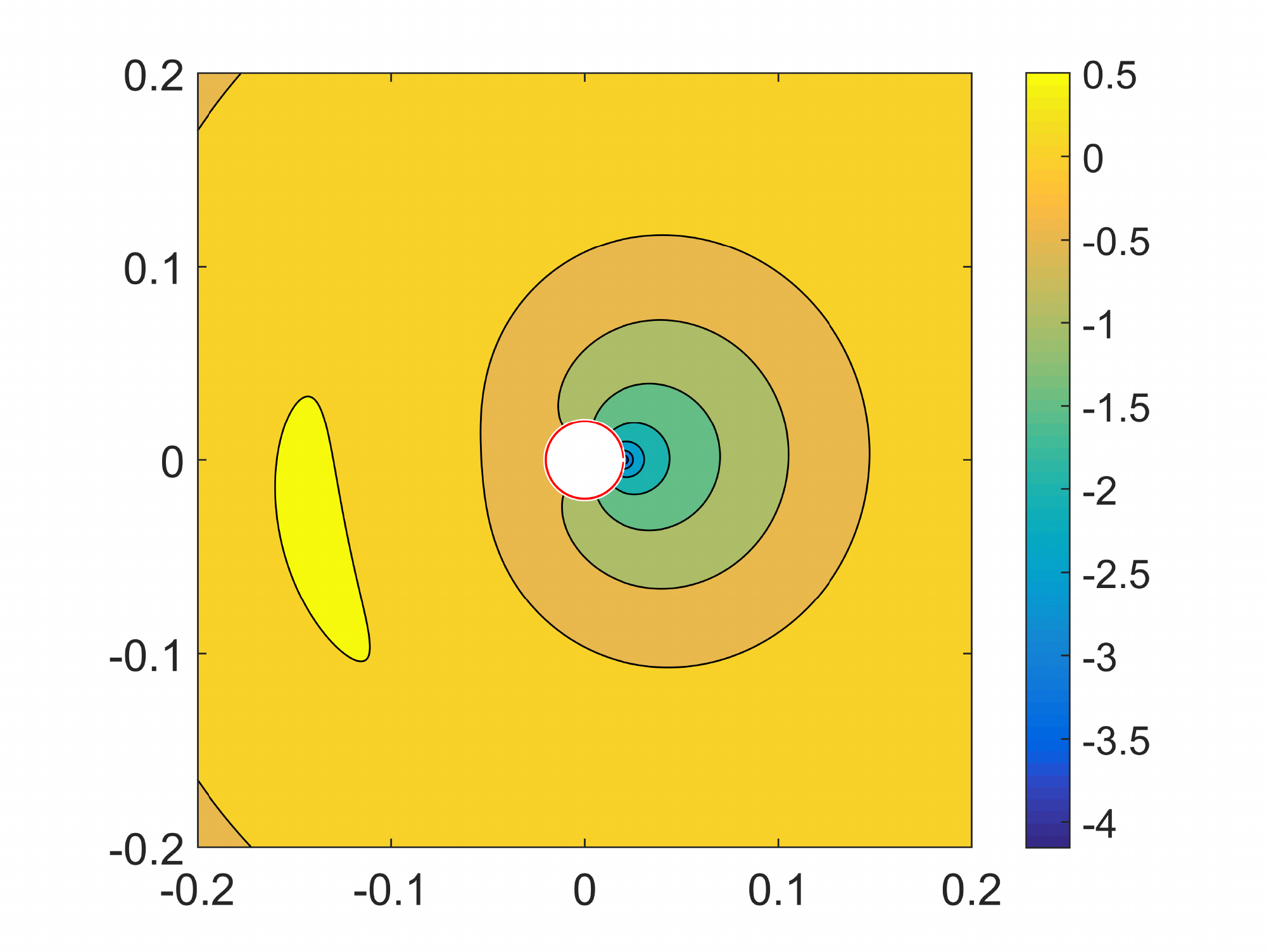}
\label{fig:figa2}}\\
\subfloat[Subfigure 2 list of figures text][]{
\includegraphics[width=0.495\textwidth]{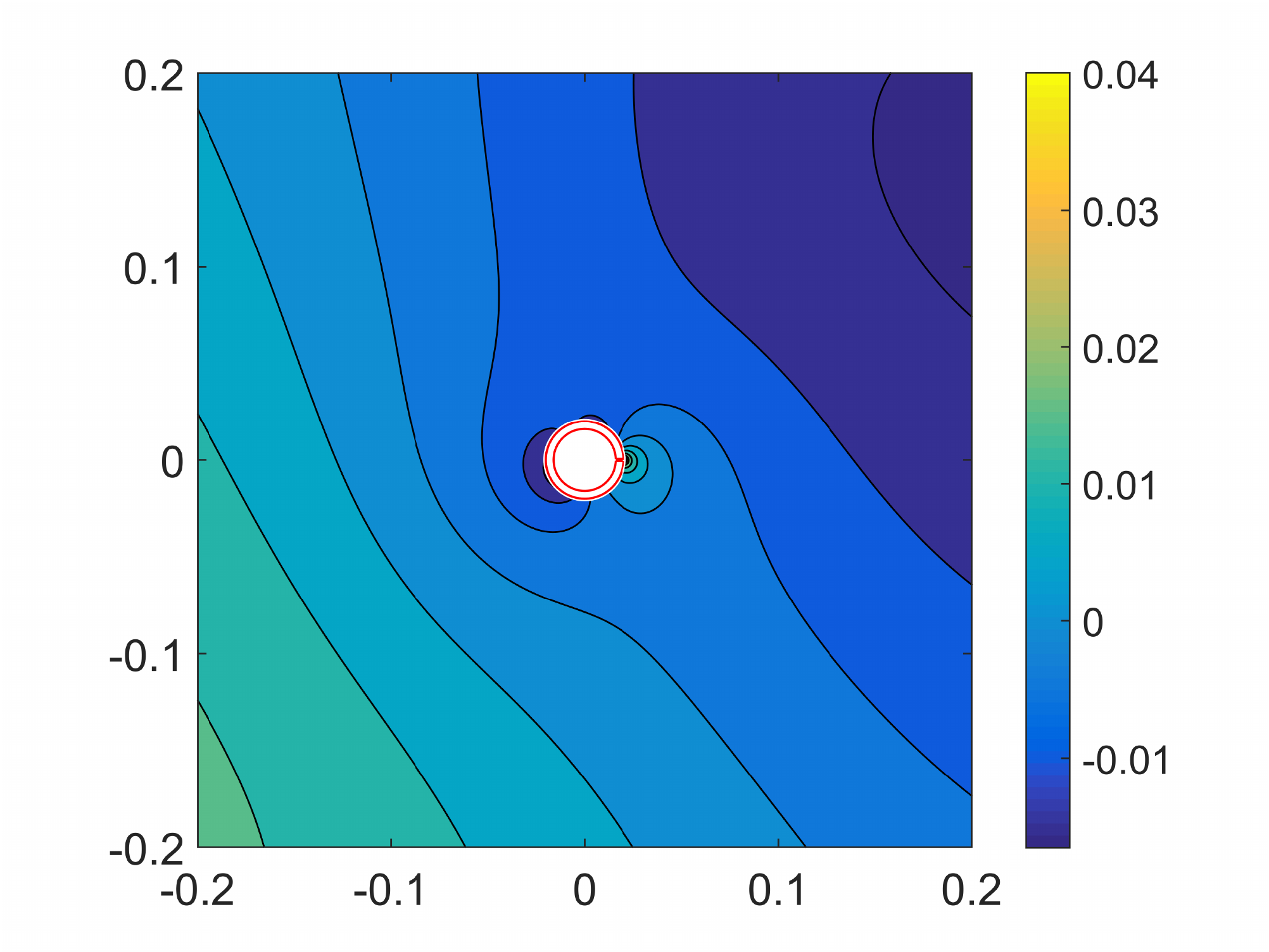}
\label{fig:figa3}}
\subfloat[Subfigure 3 list of figures text][]{
\includegraphics[width=0.495\textwidth]{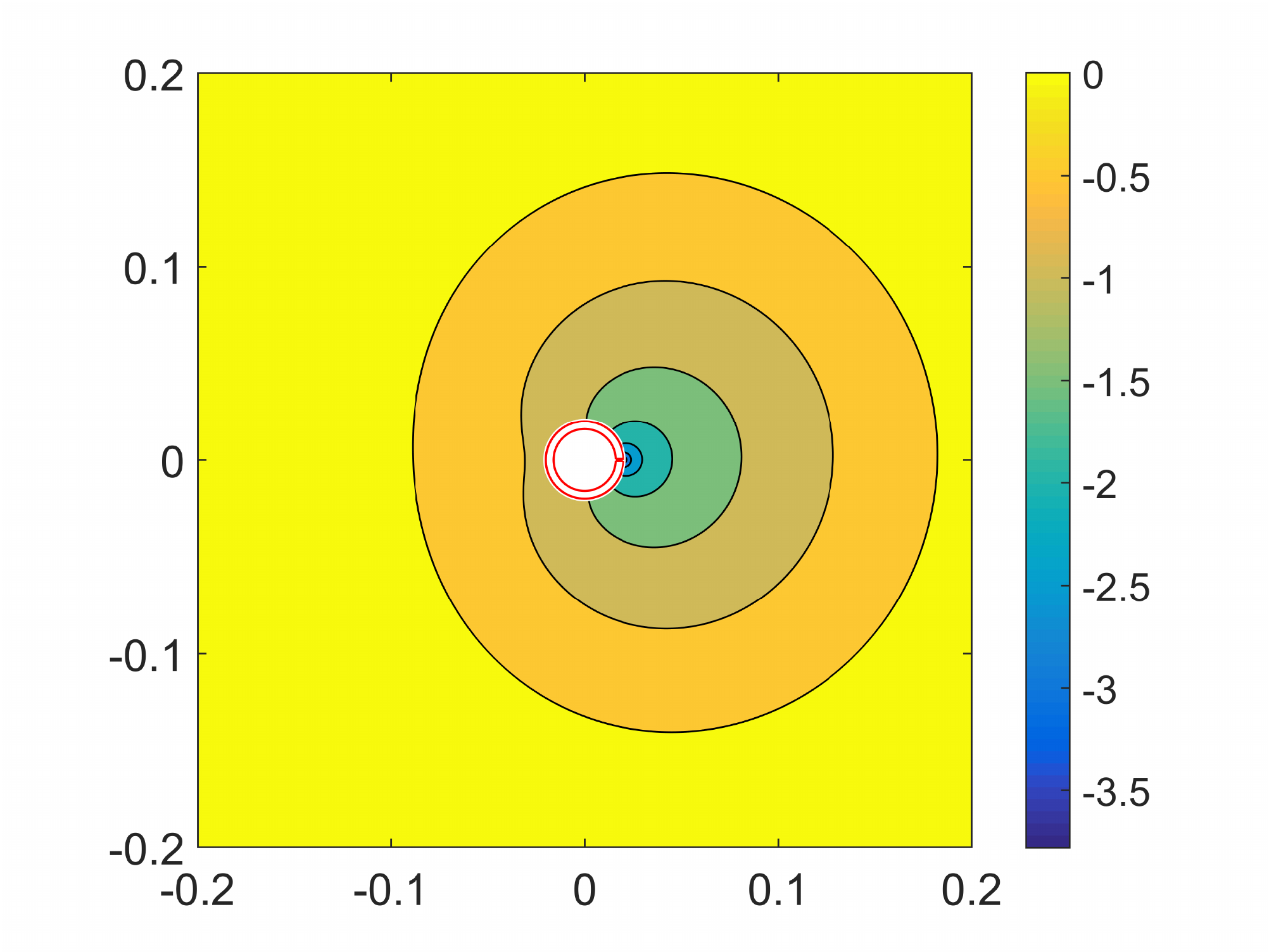}
\label{fig:figa4}}
\\
\subfloat[Subfigure 2 list of figures text][]{
\includegraphics[width=0.495\textwidth]{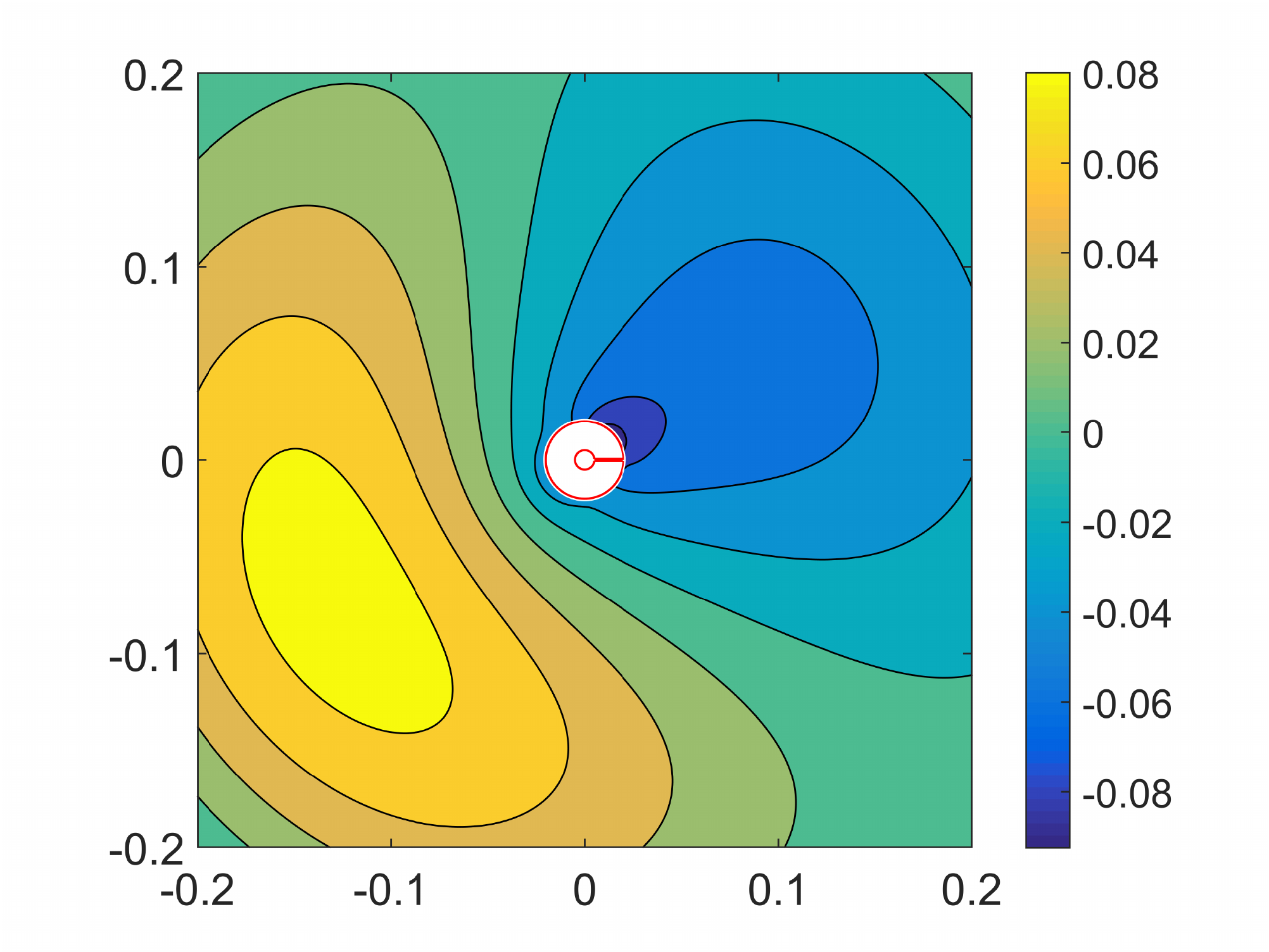}
\label{fig:figa5}}
\subfloat[Subfigure 3 list of figures text][]{
\includegraphics[width=0.495\textwidth]{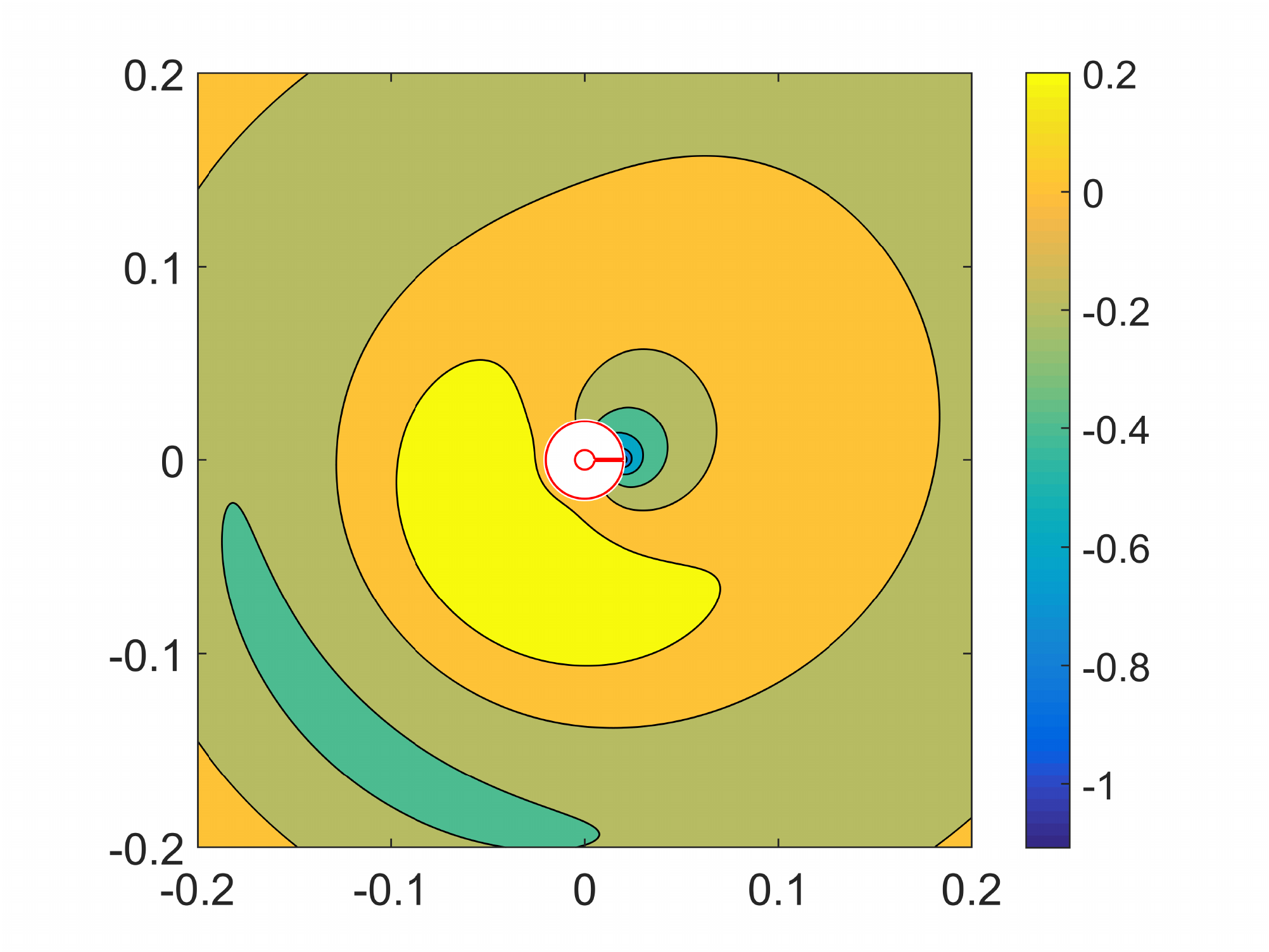}
\label{fig:figa6}}

\caption{Scattered field profiles  $\phi_\mathrm{sc}$   \eqref{eq:ansatz1} for  \protect\subref{fig:figa1},\protect\subref{fig:figa2} a thin-walled resonator at $k = k_\rmH/2$ and $k=k_\rmH \approx 16.2136$ m$^{-1}$ respectively, where $k_\rmH$ satisfies the Helmholtz resonance condition \eqref{eq:helmresthin}; \protect\subref{fig:figa3},\protect\subref{fig:figa4} a moderately thick-walled resonator ($h= 2\bar{\emm}/2\bar{\ell} = 4$) at $k_\rmH/2$ and $k_\rmH \approx 13.3001$ m$^{-1}$ respectively, where $k_\rmH$ satisfies   \eqref{eq:helmresmodthick}; \protect\subref{fig:figa5},\protect\subref{fig:figa6} a very thick-walled resonator ($h=15$) at $k_\rmH/2$ and $k_\rmH \approx 26.9376$ m$^{-1}$ respectively, where $k_\rmH$ satisfies   \eqref{eq:helmdispeqvthick}. In all figures we use   $\bar{b} = 20$\, mm, $2\bar{\ell} = 1$\,mm, and $\theta_\inc = \pi/6$.}
\label{fig:phiscatt}
\end{figure}

In Figure \ref{fig:phiscatt} we present the scattered field  $\phi_\mathrm{sc}$ for  a representative thin-walled, moderately thick-walled, and very thick-walled resonator, where all resonators possess the same outer radius $\bar{b}$ and aperture width $2\bar{\ell}$. In each setting, we examine the scattered response at  the first Helmholtz resonance frequency, $k_\rmH$,  satisfying the relevant condition in \eqref{eq:allhelmresexpr}, and at the  low frequency $k_\rmH/2$ (i.e., away from the     resonance frequency), to consider how  the field profile is modified as we approach the resonance.  In the thin-walled case we observe a field enhancement of over two orders of magnitude in the transition from $k = k_\rmH/2$ to $k=k_\rmH$, with   $\phi_\mathrm{sc}$ taking a maximum value  at the resonator mouth. A similar behaviour is observed for the moderately thick-walled case ($h=   4$), although we find that the field enhancement is approximately halved. This reduction continues for the very thick-walled resonator where the field enhancement is now slightly over one order of magnitude, and is likely due to the fact that the volume inside the resonator decreases, since we increase the aperture neck length $2\bar{\emm}$ whilst keeping the outer radius $\bar{b}$ constant. Accordingly, to achieve the strongest field enhancements  we advise that the resonator wall thickness be taken as thin as   possible. Additionally, the strongest backscattering is observed in the thin and very thick-walled representative  configurations. For reference, a  plane wave of incidence angle   $\theta_\inc =\pi/6$ is considered in all cases.

\begin{figure}[t]
\centering
\subfloat[Subfigure 6 list of figures text][]{
\includegraphics[width=0.495\textwidth]{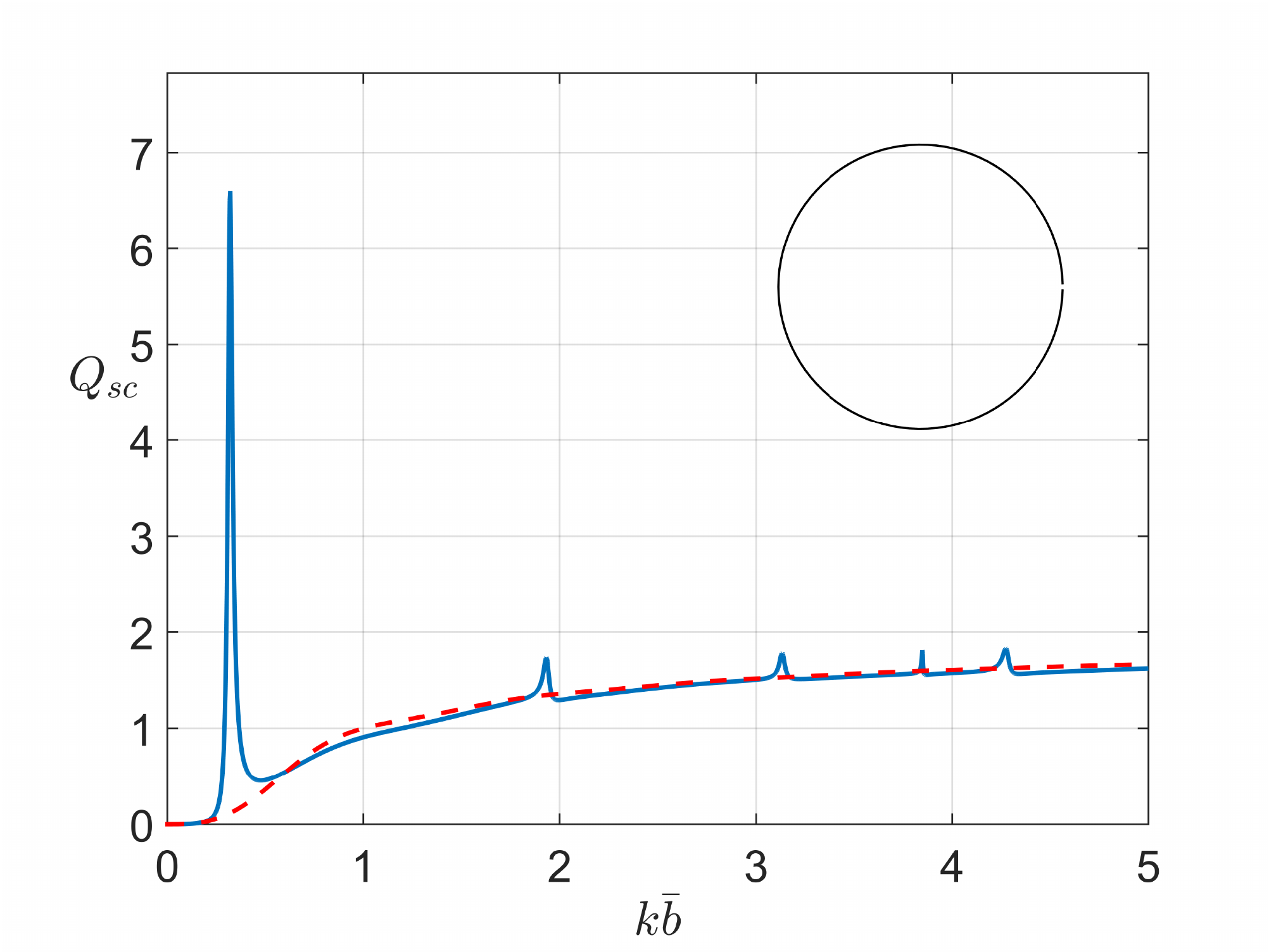}
\label{fig:figb1}}
\subfloat[Subfigure 1 list of figures text][]{
\includegraphics[width=0.495\textwidth]{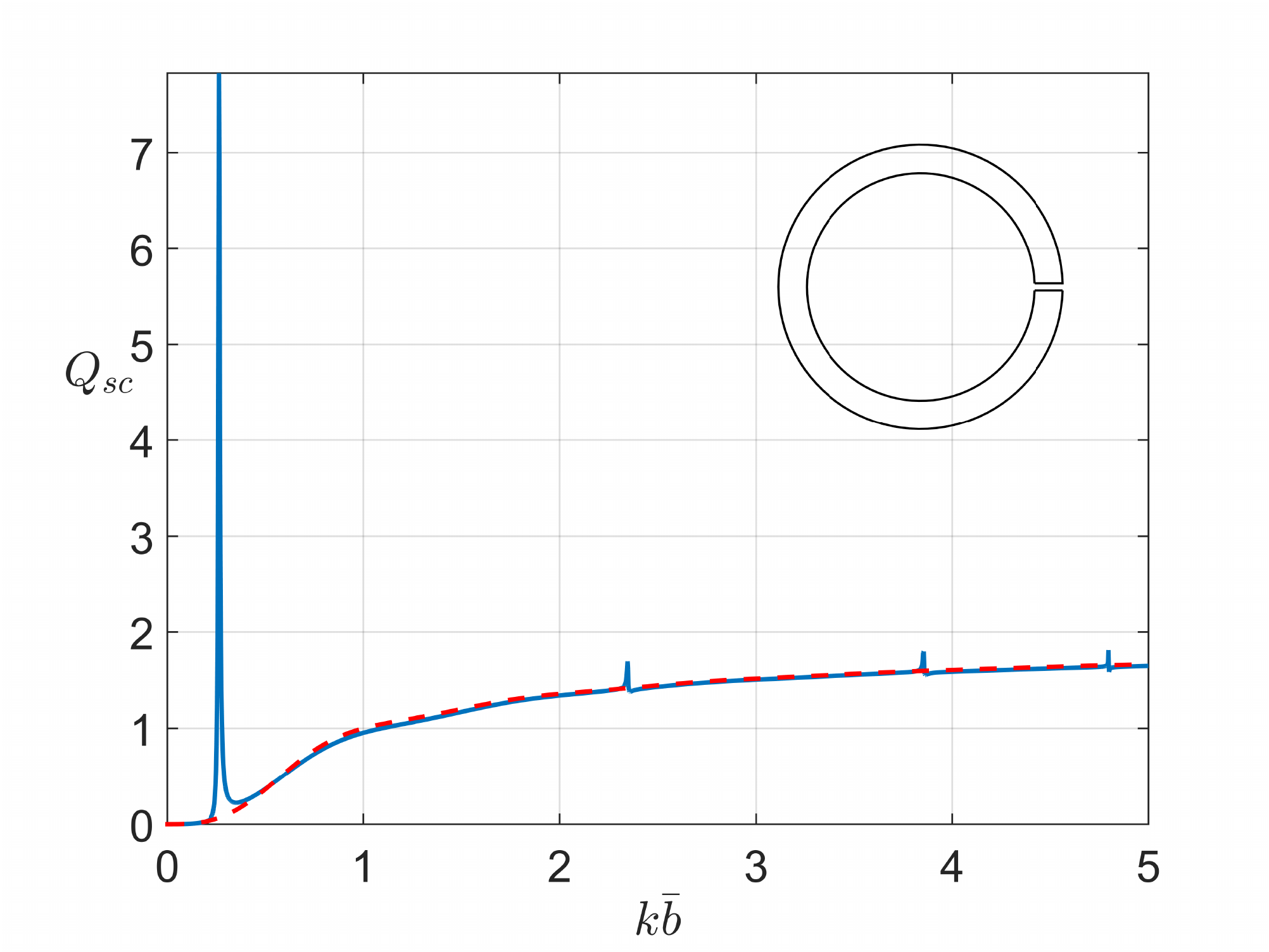}
\label{fig:figb2}}\\
\subfloat[Subfigure 2 list of figures text][]{
\includegraphics[width=0.495\textwidth]{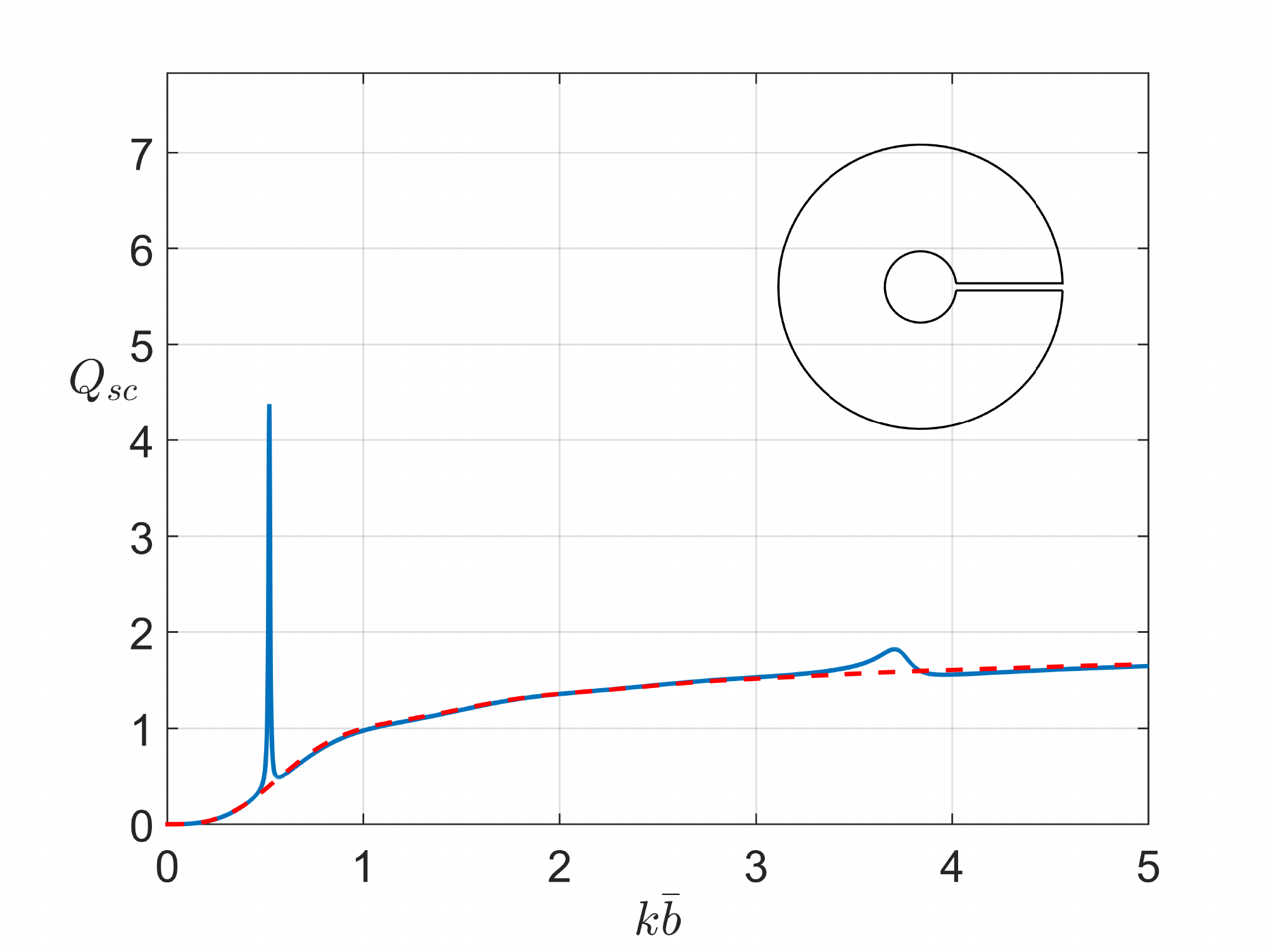}
\label{fig:figb3}}
\subfloat[Subfigure 3 list of figures text][]{
\includegraphics[width=0.495\textwidth]{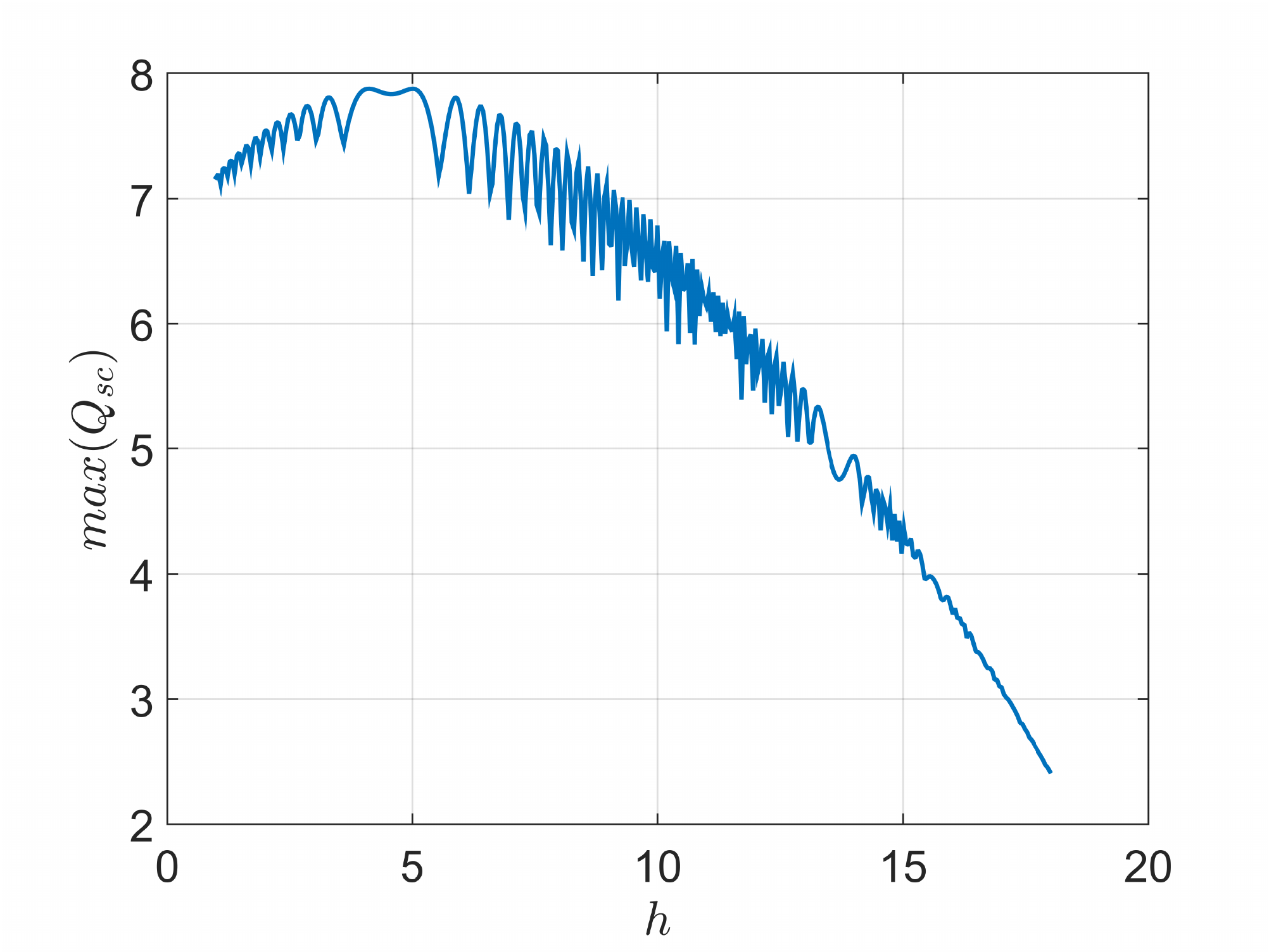}
\label{fig:figb4}}

\caption{Scattering efficiency $Q_\mathrm{sc}$ (from   \eqref{eq:sigmabarscatdef}) versus (scaled) frequency  for  a \protect\subref{fig:figb1}  thin-walled resonator, \protect\subref{fig:figb2}   moderately thick-walled resonator ($h=4$), and \protect\subref{fig:figb3}  very thick-walled resonator ($h=15$);  \protect\subref{fig:figb4}   gives the maximum value for $Q_\mathrm{sc}$ at the first Helmholtz resonance versus the channel aspect ratio $h=2\bar{\emm}/2\bar{\ell}$ (with results obtained using the thick-walled formulation).   In all figures we use   $\bar{b} = 20$ mm, $2\bar{\ell} = 1$ mm, and $\theta_\inc = \pi/6$, and results for a single closed Neumann cylinder of the same radius are superposed for reference (dashed red). Inset figures: resonator geometry (not to scale).}
\label{fig:Qscatt}
\end{figure}

In Figure \ref{fig:Qscatt} we compute the scattering efficiency $Q_\mathrm{sc}$ (following the definition for the scattering cross section $\bar{\sigma}_\mathrm{sc}$ in \eqref{eq:sigmabarscatdef}) for the three resonator configurations considered in Figure \ref{fig:phiscatt}. For all  $Q_\mathrm{sc}$ figures we superpose the result for a closed Neumann cylinder as a means of reference (dashed red curves). We find that for all resonator configurations,  a considerable enhancement is observed in the scattering efficiency  at the first Helmholtz resonance frequency $k_\rmH$ (obtained by solving \eqref{eq:allhelmresexpr} for each geometry), and  also observe that the $Q_\mathrm{sc}$ curves tend to that of a closed Neumann cylinder away from the higher-frequency Helmholtz resonance peaks. As the wall thickness increases, we observe that the spacing between Helmholtz resonances increases (as the enclosed internal resonator area becomes smaller), however we also find that the peak scattering efficiency $\max(Q_\mathrm{sc})$ does not exhibit an obvious trend. To investigate this behaviour, we  plot $\max(Q_\mathrm{sc})$ against the channel aspect ratio $h$ in Figure \ref{fig:figb4}, where we find that a maximum scattering efficiency, for our chosen outer radius and aperture width, occurs at $h \approx 4.11$ and $h\approx5.01$. The  oscillations in the $\max(Q_\mathrm{sc})$ curve are due to phase cancellation effects within the neck (see   $\phi_\mathrm{neck}$ in \eqref{eq:phineck} with $\alpha=0$), where the outward and inward propagating wave components of the solution either destructively or constructively interfere with each other.

\begin{figure}[t]
\centering
\subfloat[Subfigure 6 list of figures text][]{
\includegraphics[width=0.495\textwidth]{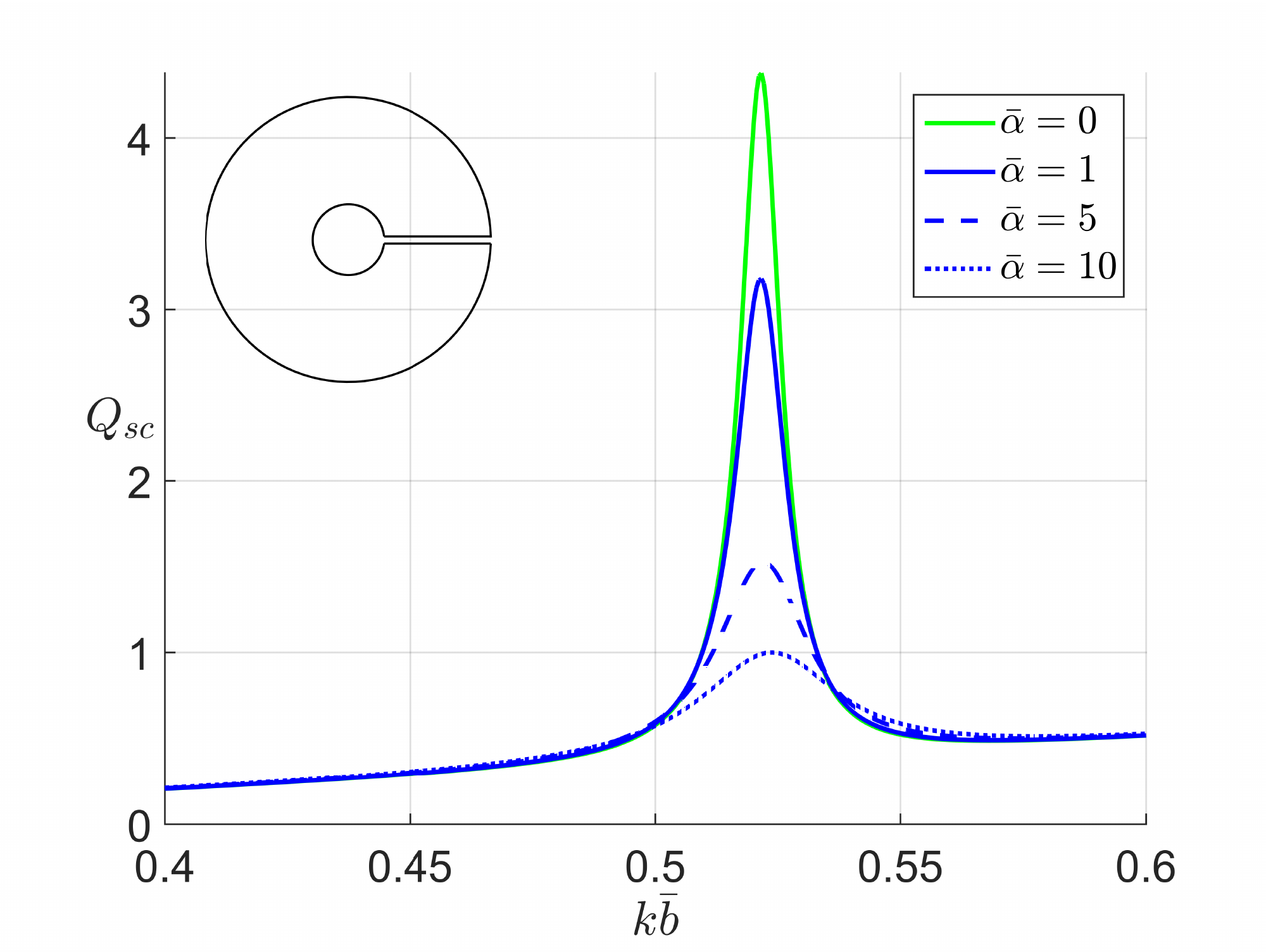}
\label{fig:figc1}}
\subfloat[Subfigure 1 list of figures text][]{
\includegraphics[width=0.495\textwidth]{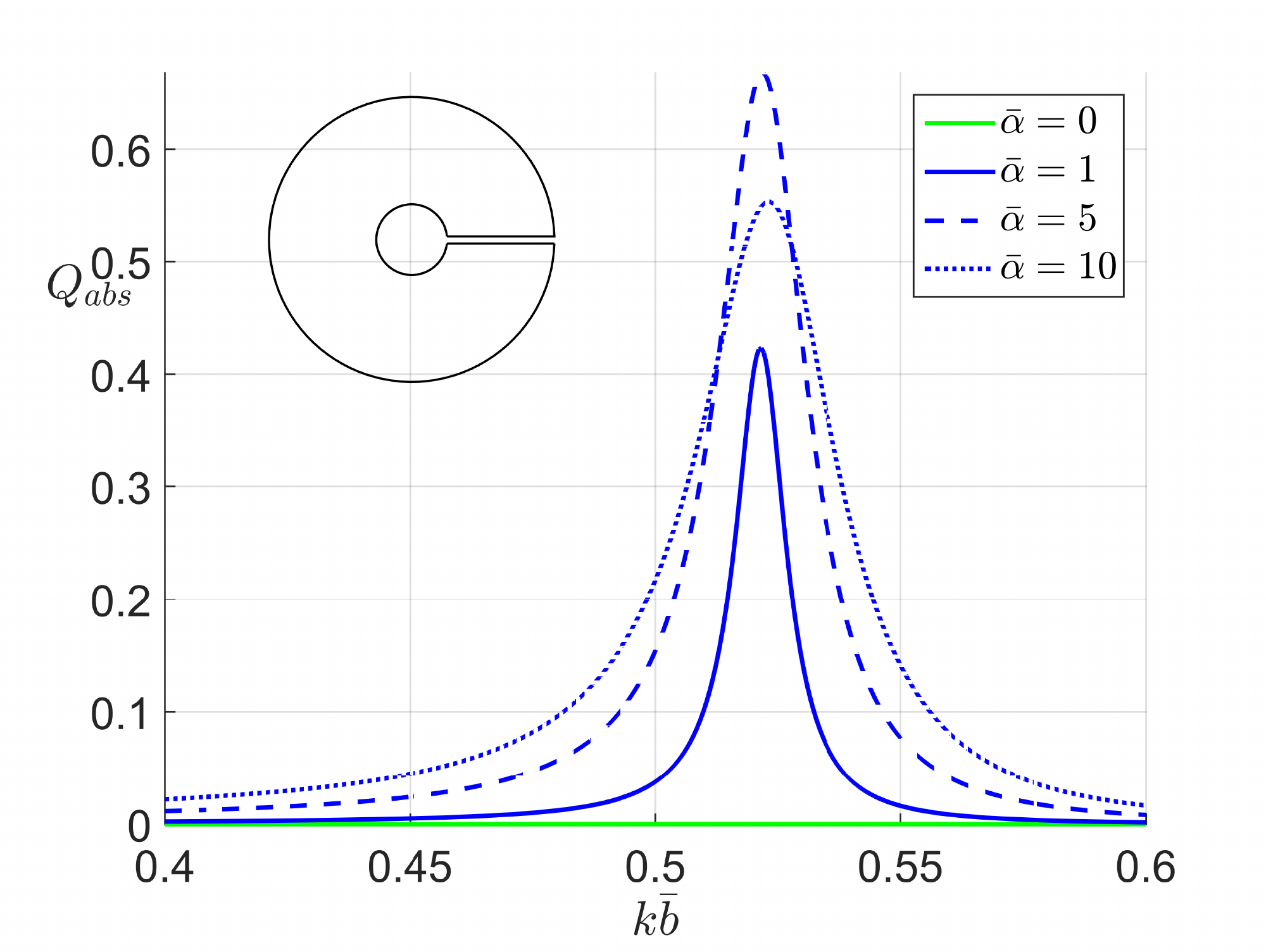}
\label{fig:figc2}}\\
\subfloat[Subfigure 2 list of figures text][]{
\includegraphics[width=0.495\textwidth]{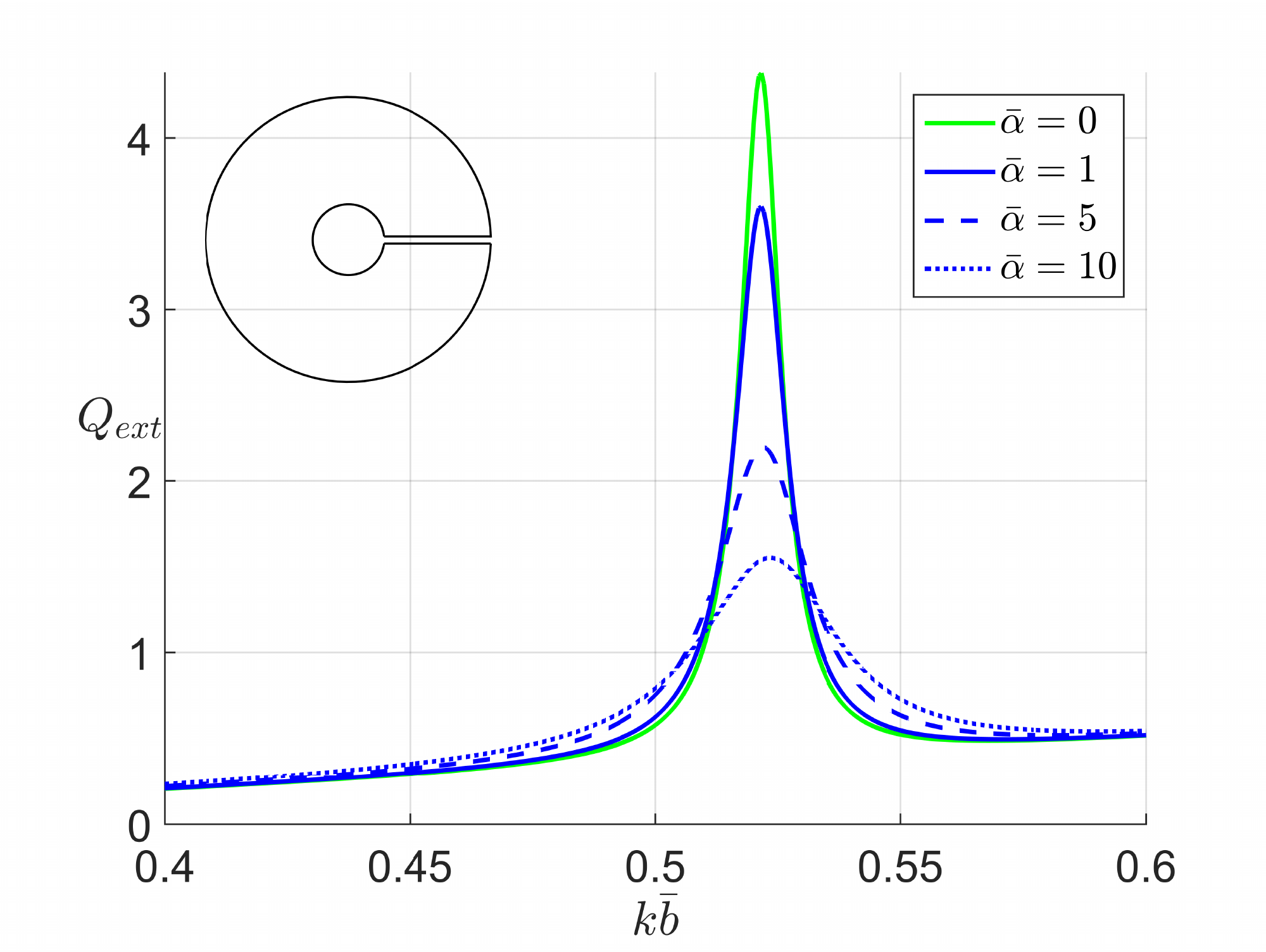}
\label{fig:figc3}}
\subfloat[Subfigure 3 list of figures text][]{
\includegraphics[width=0.495\textwidth]{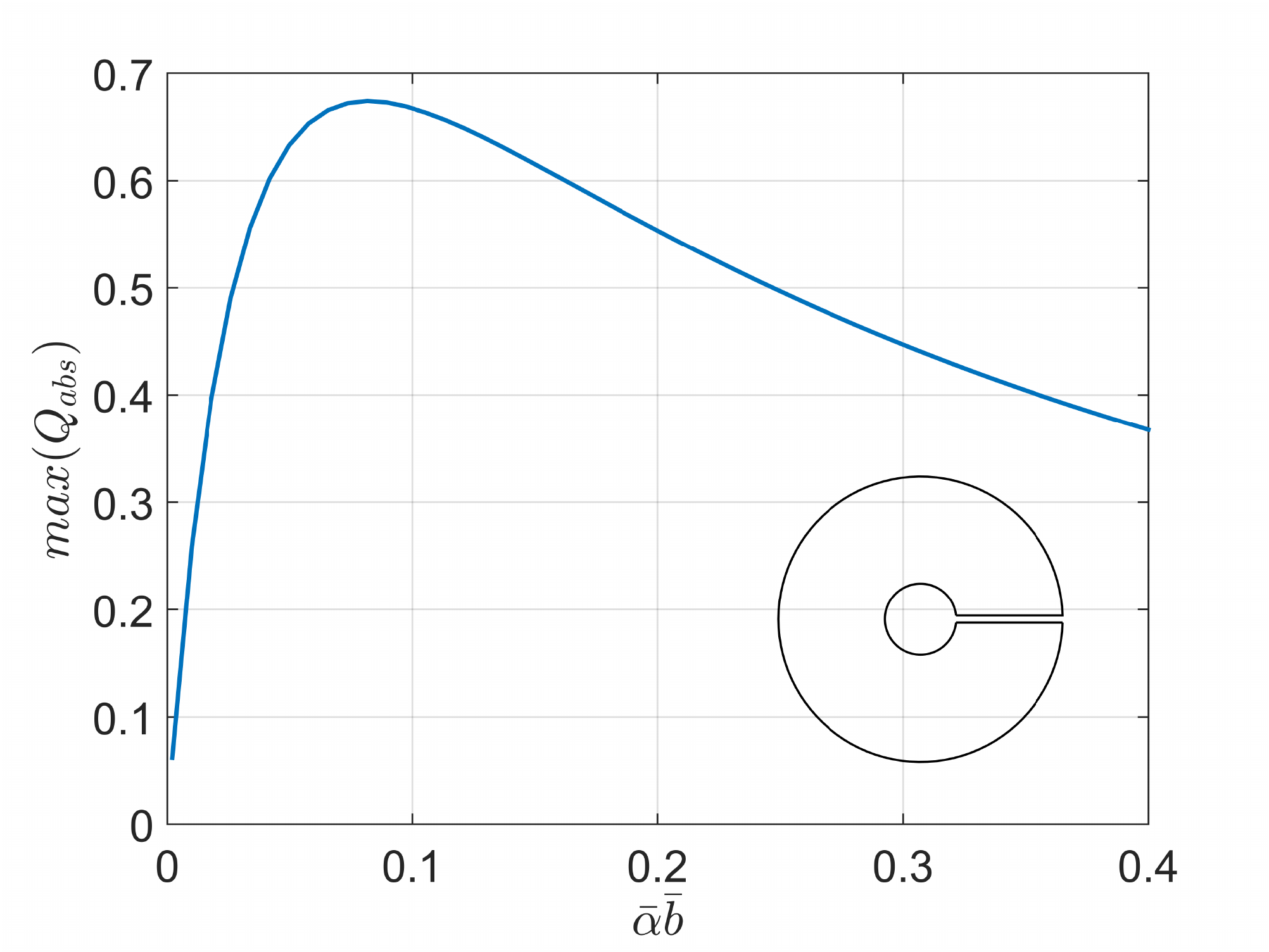}
\label{fig:figc4}}

\caption{The impact of introducing dissipation  in the neck region of a very thick-walled resonator,  via the attenuation coefficient  $\bar{\alpha}$:    \protect\subref{fig:figc1}   scattering efficiency    $Q_\mathrm{sc}$ (defined from  \eqref{eq:sigmabarscatdef}), \protect\subref{fig:figc2}    absorption efficiency $Q_\mathrm{abs}$ (defined from    \eqref{eq:sumQs}), and \protect\subref{fig:figc3}   extinction efficiency $Q_\mathrm{ext}$ (defined from \eqref{eq:sigmaextbar}), versus the (scaled) frequency;  \protect\subref{fig:figc4}   gives the maximum value for $Q_\mathrm{abs}$ at the first Helmholtz resonance versus the (scaled) attenuation coefficient  $\bar{\alpha}$.   In all figures we use   $\bar{b} = 20$\, mm, $2\bar{\ell} = 1$\,mm, $h=15$, and $\theta_\inc = \pi/6$. Inset figures: resonator geometry (not to scale).}
\label{fig:dissipneckQ}
\end{figure}

In Figure \ref{fig:dissipneckQ} we examine the impact of introducing a boundary layer in the neck of our representative very thick-walled resonator ($h=15$) from Figure   \ref{fig:phiscatt} near the first Helmholtz resonance. We compute the scattering, absorption, and extinction efficiencies for a range of attenuation values $\bar{\alpha}$, observing that in the limit as   $\bar{\alpha} \rightarrow \infty$,     results for the scattering and extinction efficiency coefficients tend to the results for (lossless) Neumann cylinders straightforwardly (see the dashed red curve in Figure  \ref{fig:figb3} for reference). Such behaviour is expected due to  the increasing resistance in the neck. However, the absorption efficiency coefficient is found to first rise  with increasing loss, and then to  decrease monotonically    towards zero; the presence of this maximum $Q_\mathrm{abs}$ therefore suggests a  range of validity   for $\bar{\alpha}$ in our treatment (and resonator geometry), i.e. $0 < \bar{\alpha}\bar{b}<0.08$ when considering dissipative loss. In general, we advise that only $\alpha$ values that lie below this peak should be considered as physically reliable. For all attenuation values we observe that  scattering  is the dominant process in the extinction of the incident power, with the reduction in $Q_\mathrm{ext}$ being driven by the reduction in $Q_\mathrm{sc}$.

\begin{figure}[t]
\centering
\subfloat[Subfigure 6 list of figures text][]{
\includegraphics[width=0.495\textwidth]{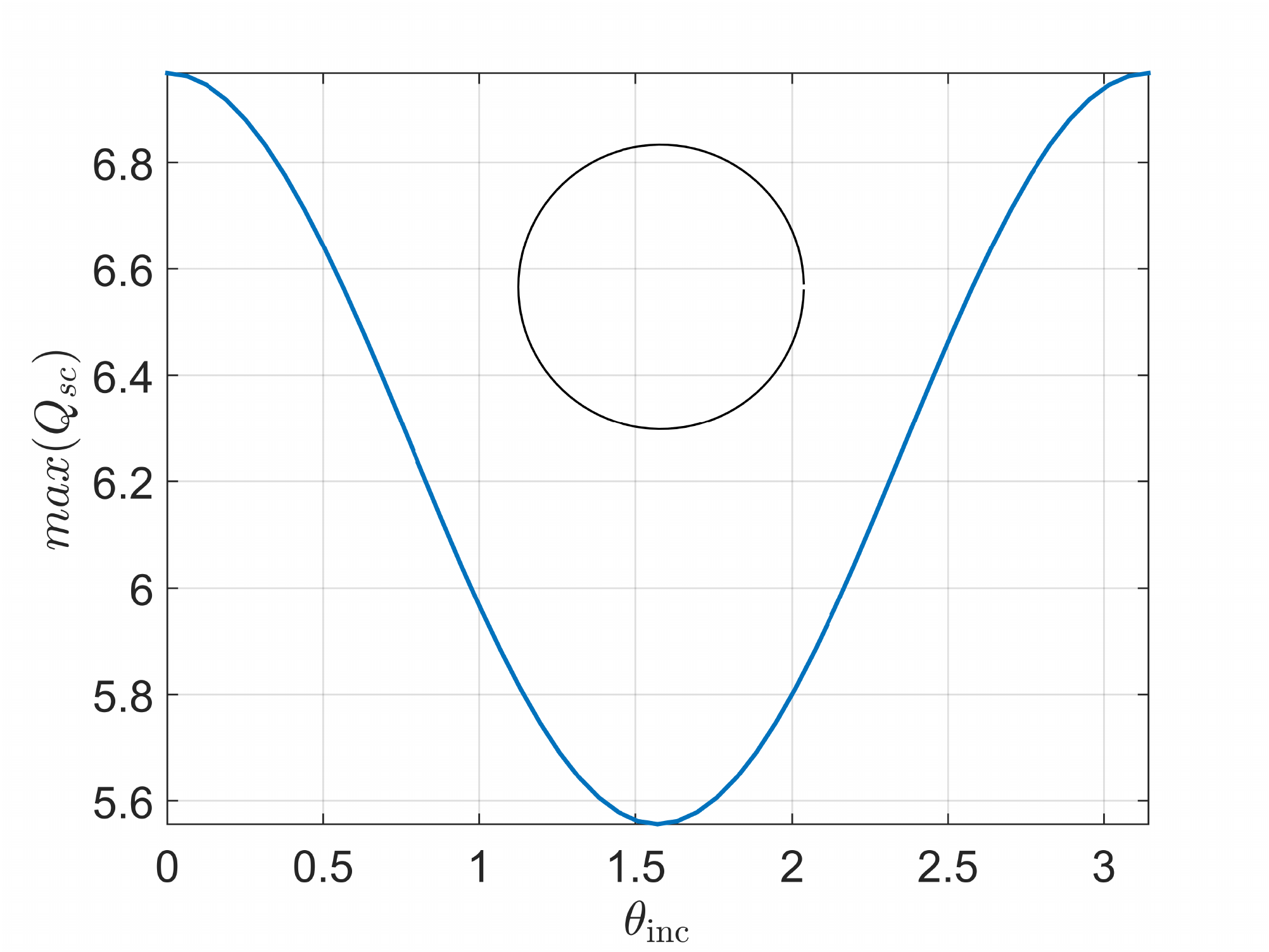}
\label{fig:figtheta1}}
\subfloat[Subfigure 1 list of figures text][]{
\includegraphics[width=0.495\textwidth]{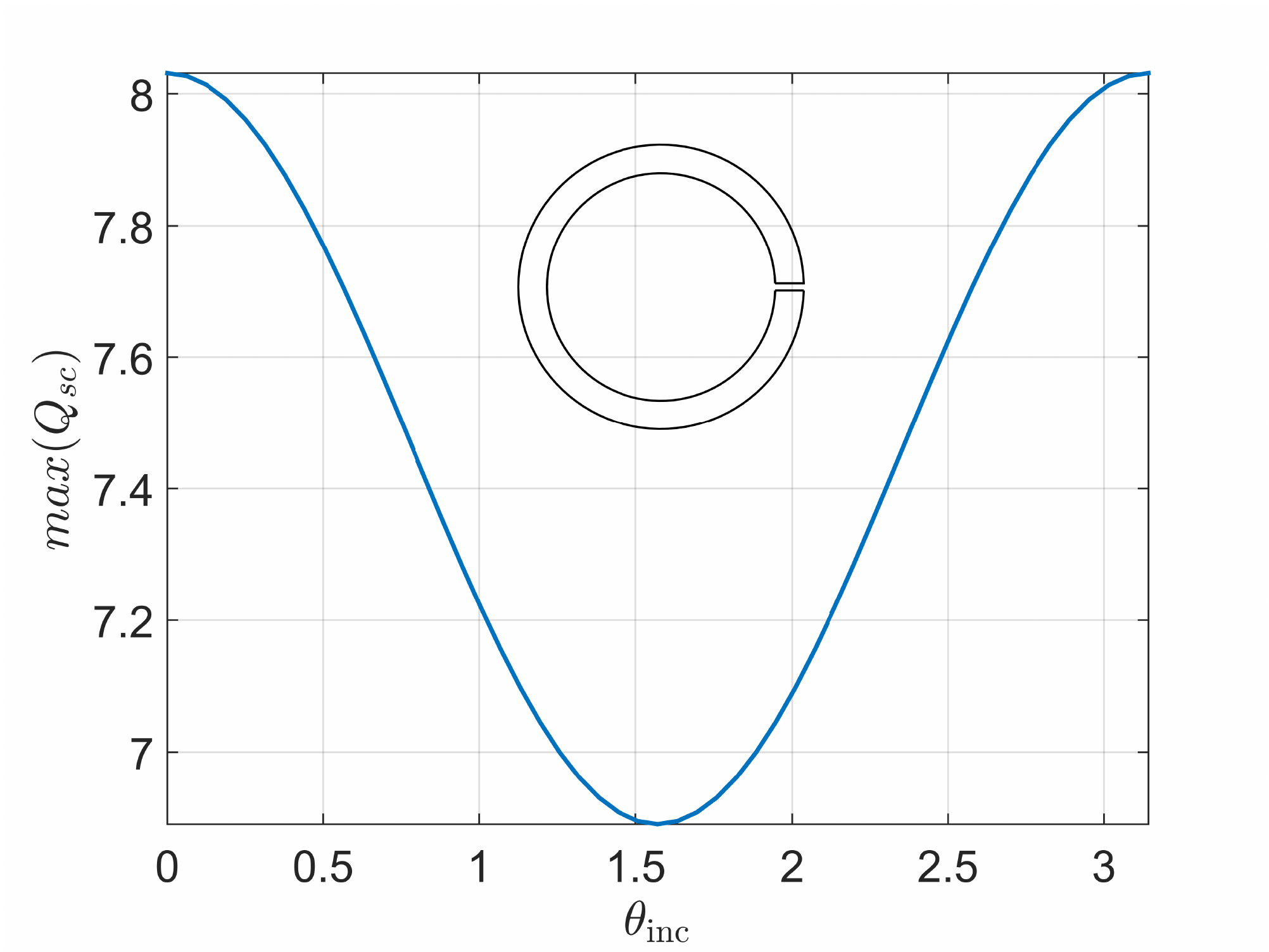}
\label{fig:figtheta2}}\\
\subfloat[Subfigure 6 list of figures text][]{
\includegraphics[width=0.495\textwidth]{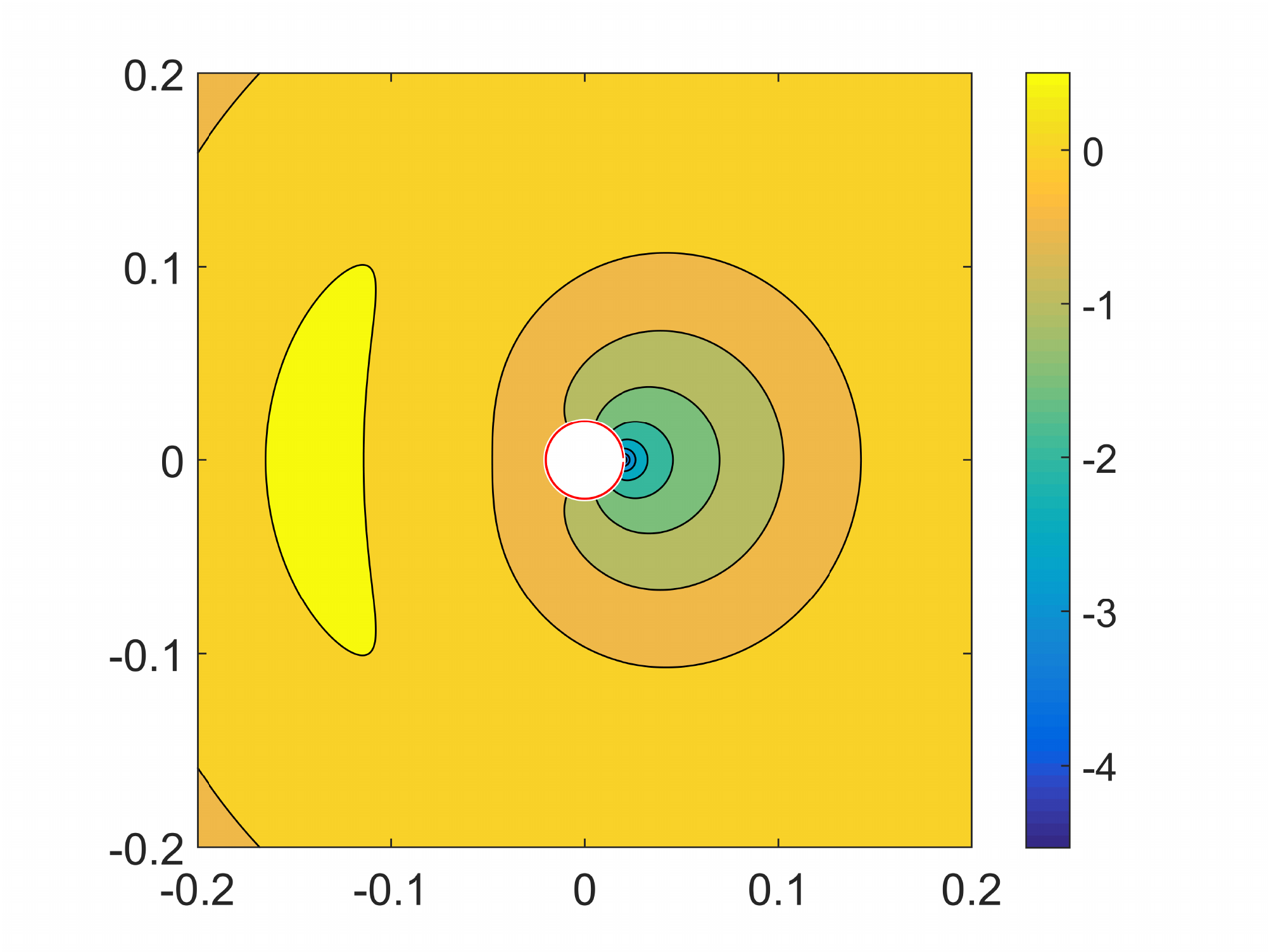}
\label{fig:figtheta3}}
\subfloat[Subfigure 1 list of figures text][]{
\includegraphics[width=0.495\textwidth]{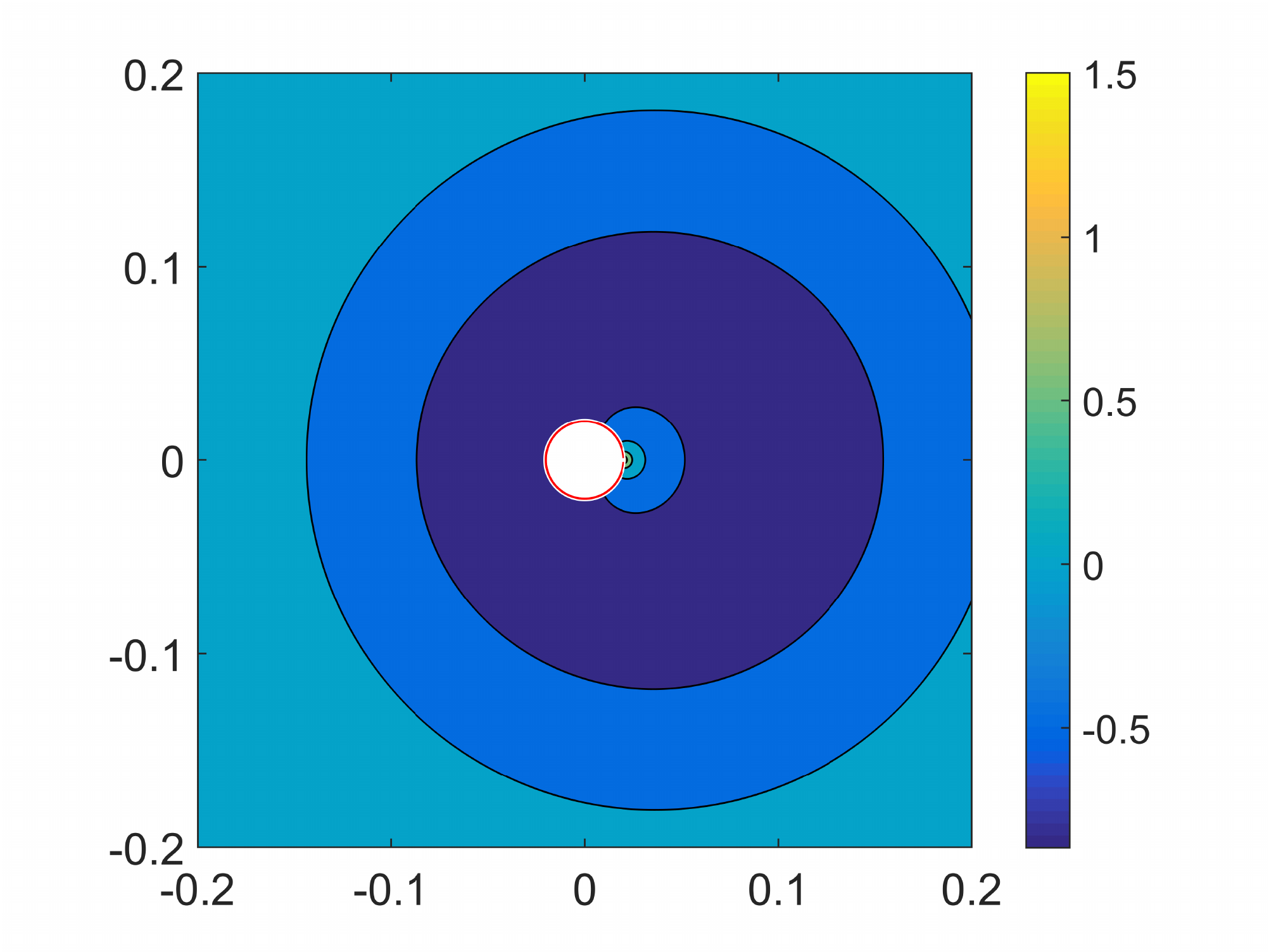}
\label{fig:figtheta4}}

\caption{The impact of   incident angle $\theta_\inc$ on  the maximum (lossless) scattering efficiency   $\max(Q_\mathrm{sc})$ (defined from \eqref{eq:sigmabarscatdef}), for a \protect\subref{fig:figtheta1}      thin-walled    and  \protect\subref{fig:figtheta2} moderately thick-walled ($h=4$) resonator.    Inset figures: resonator geometry (not to scale). Corresponding scattered field profiles $\phi_\mathrm{sc}$  \eqref{eq:ansatz1} for the thin-walled resonator configuration in Figure \protect\ref{fig:figtheta1} at $k=k_\rmH = 16.2136$ m$^{-1}$ for \protect\subref{fig:figtheta3} $\theta_\inc = 0$ and \protect\subref{fig:figtheta4} $\theta_\inc = \pi$. Here we use   $\bar{b} = 20$\,mm and $2\bar{\ell} = 1$\,mm  where applicable.}
\label{fig:sweepthetainc}
\end{figure}

 In Figure \ref{fig:sweepthetainc} we examine how the scattering efficiency $Q_\mathrm{sc}$ at the first Helmholtz resonance is impacted by the incident angle of the incoming plane wave. We present results for a thin- and moderately thick-walled resonator ($h=4$) showing that an absolute maximum scattering efficiency is observed at incidence angles along the mirror plane for the resonator (recall that the aperture is located at $\theta_0 = 0$), with a minimum observed in the orthogonal direction at $\theta_\inc = \pi/2$. That is, the incident wave does not have to be directed into the resonator $\theta_\inc=\pi$ in order to achieve maximal scattering efficiency, as the same result is obtained for $\theta_\inc = 0$.  An identical behaviour is observed for very thick resonator configurations and so we do not present the corresponding figure here. Figures \ref{fig:figtheta3} and \ref{fig:figtheta4} present the scattered field $\phi_\mathrm{sc}$ for the thin-walled resonator at the Helmholtz resonance frequency $k=k_\rmH = 16.2136$ m$^{-1}$ for the incidence angles $\theta_\inc = 0$ and $\theta_\inc=\pi$, respectively, demonstrating how very different field profiles can still return identical scattering cross sections (i.e., an identical amount of incident power scattered).

\begin{figure}[t]
\centering
\subfloat[Subfigure 6 list of figures text][]{
\includegraphics[width=0.495\textwidth]{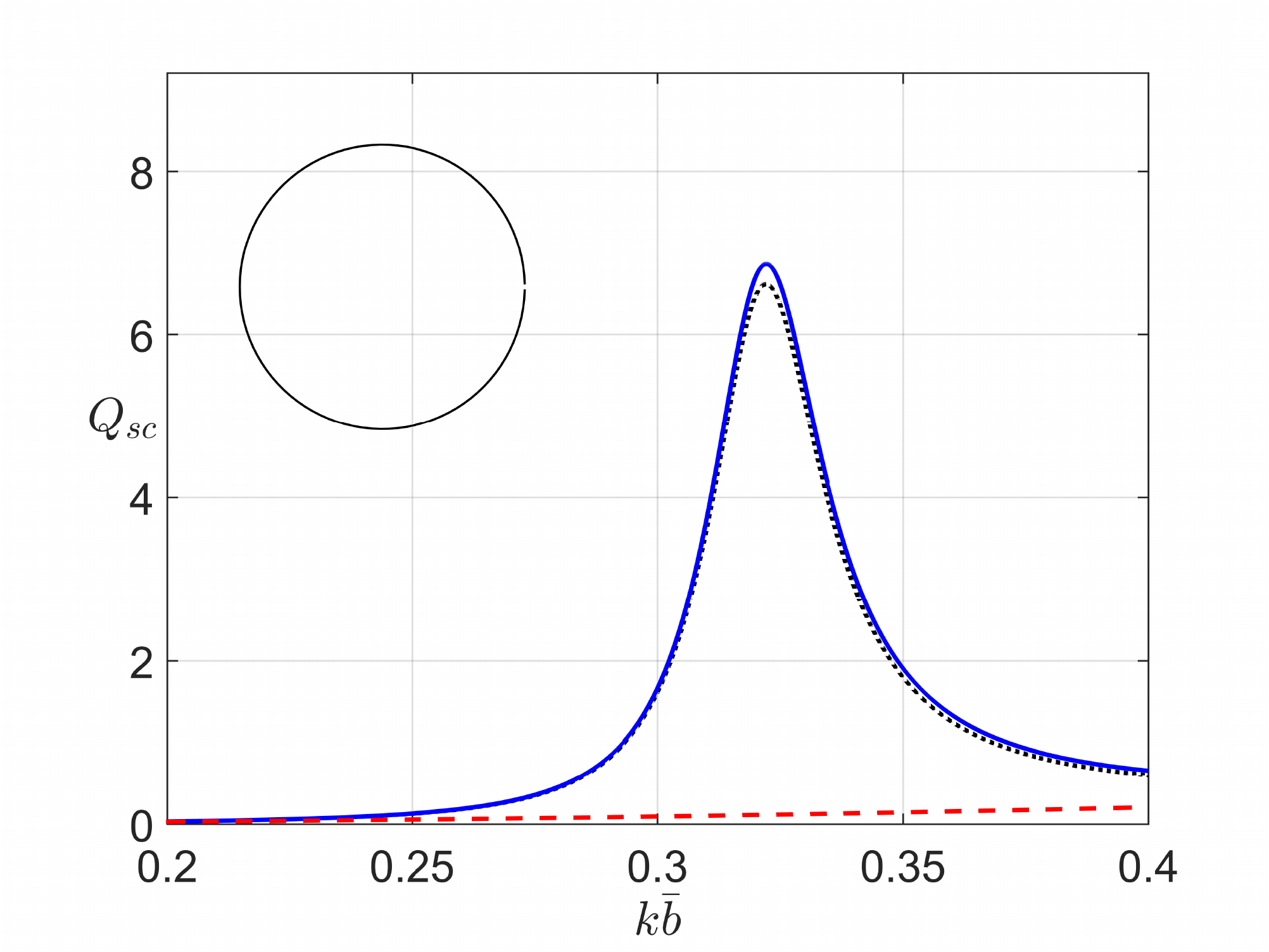}
\label{fig:figw1}}
\subfloat[Subfigure 1 list of figures text][]{
\includegraphics[width=0.495\textwidth]{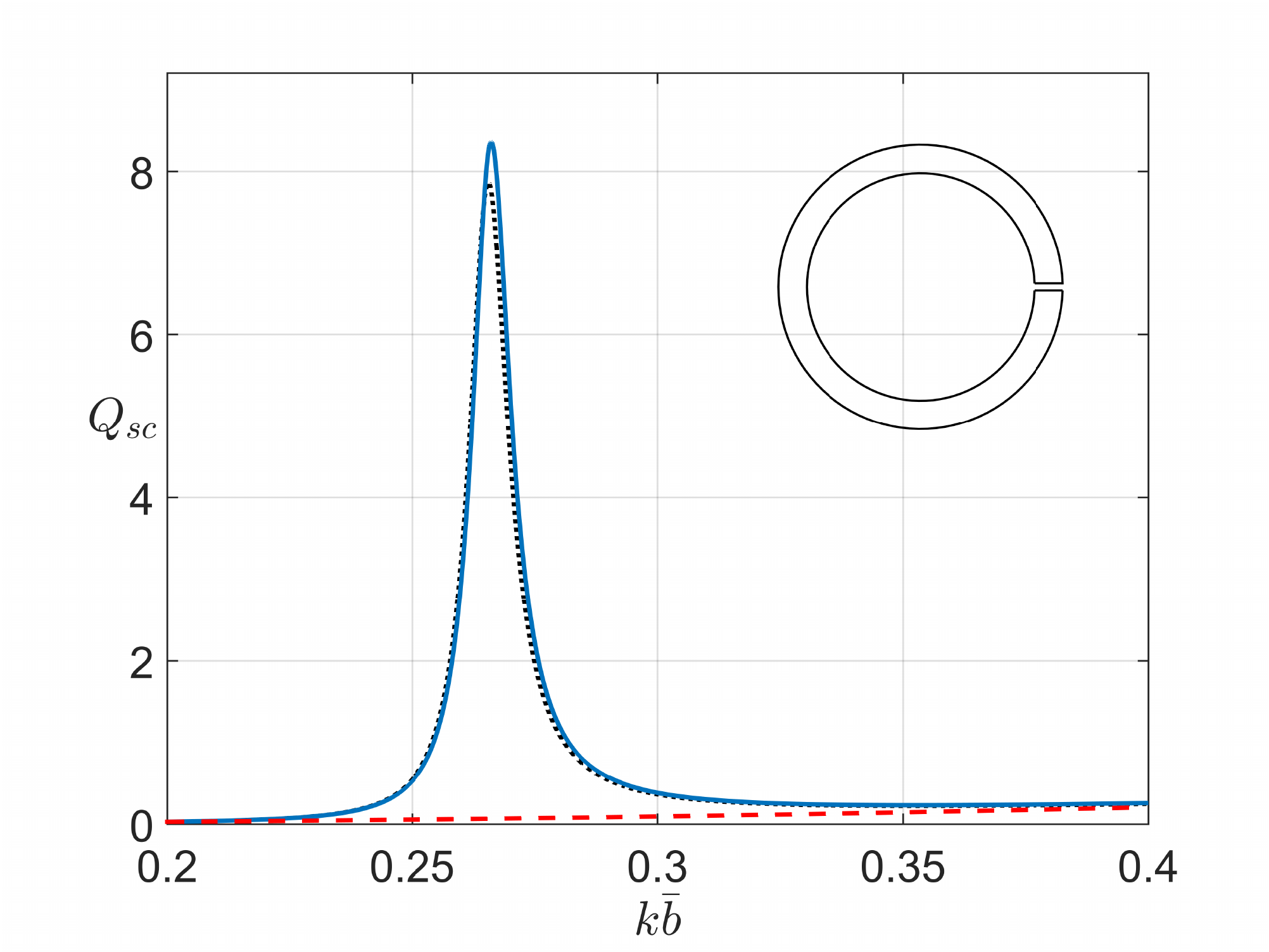}
\label{fig:figw2}}

\caption{Comparing  the scattering efficiency   $Q_\mathrm{sc}$ (defined from \eqref{eq:sigmabarscatdef}), evaluated  using the full \eqref{eq:bnfinale} (dotted black line) and asymptotic forms \eqref{eq:asyformsAqns} (blue line) for $d_n $  corresponding to   \protect\subref{fig:figw1}  a   thin-walled resonator, with the asymptotic form for $h_\varepsilon$ \eqref{eq:allbigheps} taken to $O(b^4)$, and  \protect\subref{fig:figw2} moderately thick-walled ($h=4$) resonator.   In these figures we use   $\bar{b} = 20$\,mm, $2\bar{\ell} = 1$\,mm, and $\theta_\inc = \pi$ where applicable; the result for a single closed Neumann cylinder of the same radius is superposed for reference (dashed red). Inset figure: resonator geometry (not to scale).}
\label{fig:asyscatts}
\end{figure}

In Figure \ref{fig:asyscatts} we consider    the full multipolar \eqref{eq:bnfinale}  and asymptotic \eqref{eq:asyformsAqns} forms  for $d_n $ when evaluating   the (lossless) scattering efficiency $Q_\mathrm{sc}$. Results are given for the thin-walled and moderately thick-walled ($h=4$) resonator geometries considered in Figure \ref{fig:phiscatt}. In general, we find that the asymptotic estimates  \eqref{eq:asyformsAqns} work    well  for frequencies below  the first Helmholtz resonance $k_\rmH$. However,  for the thin-walled geometry, asymptotic forms for $h_\varepsilon$ \eqref{eq:allbigheps} must be taken to $O(b^4)$ in order to recover the $Q_\mathrm{sc}$ peak as shown; this representation of $h_\varepsilon$ is not presented here for compactness. Additionally, results for the very thick-walled ($h=15$) configuration are not presented here as we require asymptotic forms for $h_\varepsilon$ and $d_{0,\pm1,\pm2} $ to a very high order in $b$ for accuracy, although   for $k \ll k_\rmH$ we find that the asymptotic forms \eqref{eq:asyformsAqns} for $d_n $ work  well   for all  resonator configurations.

For reference, numerical investigations examining  the impact of varying the channel aspect ratio $h$ and the aperture half-angle $\theta_\mathrm{ap}$ on the first Helmholtz resonance frequency $k_\rmH$ for a very thick-walled resonator ($h=15$) return a similar qualitative behaviour to that seen in Smith and Abrahams \cite{smith2022PartII} for two-dimensional arrays. Additionally,   the   cross section results for lossless cylinders match those presented in Rubinow \cite{rubinow1956first}  after a factor 2 correction in their  definition  is taken into account.

\section{Discussion} \label{sec:disc}
In this paper we have presented an analytic solution method for deducing  plane wave scattering by   a single two-dimensional Helmholtz resonator. Our solution procedure,     valid  at low frequencies,  combines multipole methods with the method of matched asymptotic expansions, extending results from earlier works by the authors on homogeneous  two-dimensional arrays of resonators \cite{smith2022PartI,smith2022PartII}. In addition to describing the scattered field, we determine the extinction, absorption, and scattering cross sections for a selection of resonator designs and compare these against results for an isolated Neumann cylinder.  

 Numerical investigations     demonstrate  considerable field enhancement near the resonator mouth, at the first Helmholtz resonance frequency, with a strong dependence on the wall thickness. We have   found that    optimal wall thicknesses exist to achieve maximal cross sections (efficiencies) for a prescribed outer radius and incidence angle. Furthermore, we consider the impact of a boundary layer emerging in the resonator neck, giving rise to viscous dissipative losses. Although a lossy neck gives rise to moderate values for the absorption efficiency, the corresponding reduction in	the scattering   efficiency     has the net effect of diminishing the extinction efficiency at the first Helmholtz resonance and beyond, where in the limit of large loss, we find that all cross sections return  to the results for a Neumann cylinder, as expected. In general the maximal scattering efficiency $Q_\mathrm{sc}$ is found for incidence wave angles that lie along the mirror symmetry plane for the resonator, despite the fact that very different scattered field profiles are observed for $\theta_\inc = 0$ and $\theta_\inc = \pi$. The formulation presented here should prove useful for ongoing theoretical and experimental work by other groups \cite{su2018retrieval,melnikov2019acoustic,quan2020nonlocality}.

Recently, Melnikov {\it et al.} \cite{melnikov2019acoustic}  has conducted experimental and theoretical work on plane wave scattering by a single Helmholtz resonator of the type considered here. In their work they use a {\it lumped-element model}   \cite{pierce2019acoustics} to describe the resonator response which takes the form of a third-order ordinary differential equation  (a non-linear   spring model). Such phenomenological-type modelling is  not required to determine the plane wave scattering response of a single resonator,   as we have shown here; that is, the   asymptotic matching procedure recovers  field representations that are theoretically indistinguishable from the genuine field at low frequencies.  Furthermore,  Melnikov {\it et al.} \cite{melnikov2019acoustic}   constructs an acoustic analogue to the electric {\it polarisability tensor} in electrostatics \cite{milton2002theory}, with non-zero off-diagonal terms   that are referred to as elements of a Willis coupling tensor    (see Appendix \ref{sec:willis} in the present work for an in-depth discussion). However, we stress that Willis coupling in general refers to the tensors that emerge within generalised constitutive relations for   effective structured media, and furthermore,  as shown  in   earlier work on two-dimensional arrays of resonators by the authors \cite{smith2022PartII}, that  anisotropy and not bianisotropy is observed in bulk. Accordingly, the presence of Willis-like behaviour (depolarisability effects)  in the  acoustic polarisability  tensor    \cite{melnikov2019acoustic}, as shown in our Appendix, does not necessarily correlate with an effective   Willis tensor effect in bulk. Although Willis coupling tensors are vanishing at low frequencies (for centrosymmetric unit cells), they are still nonetheless present at low frequencies, and given their absence in the effective dispersion equation in bulk \cite{smith2022PartI,smith2022PartII}, we therefore see no evidence that two-dimensional arrays of Helmholtz resonators  exhibit bianisotropy (Willis coupling).

In addition to how this work may relate to Willis coupling, future work   includes  careful parameter sweeps to examine the impact of aperture width on all cross sections, as well as     studies considering scattering by random suspensions of resonators, finite clusters, and one-dimensional arrays (with the latter currently under investigation by the authors). The impact of different internal resonator geometries (i.e., square) may be of interest. Although the results derived here are formally derived for small aperture widths $2\bar{\ell}$, our treatment appears to hold for much wider apertures than that considered numerically in the present work (see \cite{smith2022PartI,smith2022PartII}). Other future work includes an extension of our model to incorporate nonlinear  effects in the fluid within the neck, which removes the need to assume the existence of nonlinearity {\it a priori}, as is required for   lumped element model treatments.

\appendix 

\section{Acoustic polarisability and Willis-like coupling} \label{sec:willis}
When designing periodically structured media  it is important to consider the   possibility that a metamaterial, comprised of purely linear phases, may    behave as a Willis (bi-anisotropic) medium. It is also possible that the metamaterial may exhibit enhanced  nonlinear constitutive behaviour \cite{smith2015electrostriction}, but we do not consider that possibility here.  Recently, a selection of works   \cite{su2018retrieval,melnikov2019acoustic,quan2020nonlocality} have claimed that an analogue   to Willis coupling is exhibited in the scattering response of a single Helmholtz resonator, which we now examine using our present formalism.

Following the conventions outlined in Quan \cite{quan2020nonlocality} we begin by writing the (dimensional) scattered pressure for the resonator in terms of an ideal (point-source) monopole and dipole at the origin
\begin{equation}
\overline{P}_\mathrm{sc} = -\rmi k^2 \frac{c_\rmp}{4}  {M} \rmH_0^{(1)}(k \overline{r}) 
-\rmi \frac{k^3 c_\rmp^2}{4} \left( {D}_{\overline{x}} \cos\theta +  {D}_{\overline{y}} \sin\theta \right)\rmH_1^{(1)}(k \overline{r}) ,
\end{equation}
where $ \overline{r}$ denotes the dimensional distance from the origin,   $c_\rmp = \sqrt{B/\rho}$ is the free-space phase velocity, $ {M}$ is the acoustic monopole amplitude,   $( {D}_{\overline{x}}, {D}_{\overline{y}})$ are the acoustic dipole amplitudes, and we consider   $\exp(-\rmi \omega t)$ time-harmonic dependence as in the main text. Through the use of orthogonality relations it follows that  \cite[Eq.(2.51)]{quan2020nonlocality}
\begin{subequations}
\label{eq:MDXDY}
\begin{align}
M &= \frac{2 \rmi c_\rmp^{-2}}{\pi k^2 \rmH_0^{(1)}(k\overline{r})} \, \int_{0}^{2\pi}\overline{P}_\mathrm{sc}(\overline{r},\theta)  \, \rmd \theta, \\
D_{\overline{x}} &= \frac{4 \rmi c_\rmp^{-2}}{\pi k^3 \rmH_1^{(1)}(k\overline{r})} \, \int_{0}^{2\pi}\overline{P}_\mathrm{sc}(\overline{r},\theta)  \cos(\theta) \, \rmd \theta, \\
D_{\overline{y}} &= \frac{4 \rmi c_\rmp^{-2}}{\pi k^3 \rmH_1^{(1)}(k\overline{r})} \, \int_{0}^{2\pi}\overline{P}_\mathrm{sc}(\overline{r},\theta)  \sin(\theta) \, \rmd \theta, 
\end{align}
\end{subequations}
where $\overline{P}_\mathrm{sc} = \rmi \omega \rho \overline{\phi}_\mathrm{sc}$ in turn follows from the linearised form of Bernoulli's law \cite{pierce2019acoustics}. Thus, using the  dimensionless form of $\phi_\mathrm{sc}$   given in \eqref{eq:expanphiext2}, we are able to write \eqref{eq:MDXDY} in the form
\begin{equation}
\label{eq:polarisabilitytensor}
\left[
\begin{array}{c}
M \\[5pt]
D_{\overline{x}}\\[5pt]
D_{\overline{y}}
\end{array}
\right]
=
\left[
\begin{array}{ccc}
 {\alpha}_{11}&  {\alpha}_{12} & 0 \\[5pt]
 {\alpha}_{21} &  {\alpha}_{22} & 0 \\[5pt]
0 & 0 &  {\alpha}_{33}
\end{array}
\right]
\left[
\begin{array}{c}
\overline{P}_\inc^{(0)}  \\[5pt]
\bar{v}_{\inc}^{(0)} \\[5pt]
\bar{w}_{\inc}^{(0)}
\end{array}
\right],
\end{equation}
where
\begin{subequations}
\label{eq:allalphaslongform}
\begin{align}
 {\alpha}_{11} &= \frac{4\rmi}{(c_\rmp k)^2} \left( \frac{2 h_\varepsilon^{-1}}{\left[\pi k\bar{b} H_0^{(1)\prime}(k\bar{b})\right]^2 } - \frac{\rmJ_0^\prime(k\bar{b})}{\rmH_0^{(1)\prime}(k\bar{b})}	 \right) , \quad
 {\alpha}_{12}  = -\frac{16\rho h_\varepsilon^{-1}}{\pi^2 c_\rmp k^4 \bar{b}^2 \rmH_0^{(1)\prime}(k\bar{b}) \rmH_1^{(1)\prime}(k\bar{b})}, \\
 {\alpha}_{21} &= \frac{16 \rmi h_\varepsilon^{-1}}{\pi^2 k^5 \bar{b}^2 c_\rmp^2  \rmH_0^{(1)\prime}(k\bar{b})\rmH_1^{(1)\prime}(k\bar{b})}, \quad
 {\alpha}_{22}   = \frac{8\rho}{k^3 c_\rmp} \frac{\rmJ_1^\prime(k\bar{b})}{\rmH_1^{(1)\prime}(k\bar{b})} - \frac{32 \rho h_\varepsilon^{-1}}{c_\rmp k^3 \left[\pi k \bar{b} \rmH_1^{(1)\prime}(k\bar{b})\right]^2}, \\
 {\alpha}_{33} &= \frac{8\rho }{k^3 c_\rmp}\frac{\rmJ_1^\prime(k\bar{b})}{\rmH_1^{(1)\prime}(k\bar{b})},
\end{align}
\end{subequations}
with $\overline{P}_\inc^{(0)} = \overline{P}_\inc(\overline{r} = 0) = \rmi k c_\rmp \rho $ denoting the incident pressure, and the vector $(\bar{v}_{\inc}^{(0)},\bar{w}_{\inc}^{(0)}) = (\partial_{\overline{x}} \overline{\phi}_\inc|_{\overline{r} = 0},\partial_{\overline{y}}\overline{\phi}_\inc|_{\overline{r} = 0})$ representing the incident velocity,  at the origin. In the limit as $b\rightarrow 0$ we find that
\begin{subequations}
\begin{align}
 {\alpha}_{11} &\sim \frac{2\rmi}{(k c_\rmp)^2 h_\varepsilon} \left[ -1 + (k\bar{b})^2 \left(\frac{1}{2} - \gamma_\rme + \frac{\rmi \pi}{2} - \log\left(\frac{ k\bar{b} }{2}\right) \right)\right] - \frac{\pi \rmi (k\bar{b})^2}{4},
\quad 
 {\alpha}_{12} \sim \frac{4\rho \bar{b}}{c_\rmp k  h_\varepsilon}, 
\\
 {\alpha}_{21} &\sim -\frac{4\rmi   \bar{b}}{c_\rmp^2 k^2 h_\varepsilon},
\quad
 {\alpha}_{22} \sim -\frac{2\pi \rmi \rho   \bar{b} ^2}{c_\rmp k } + \frac{8\rho   \bar{b} ^2}{c_\rmp k  h_\varepsilon},
\quad
 {\alpha}_{33} \sim -\frac{2\pi \rmi \rho   \bar{b} ^2}{c_\rmp k }.
\end{align}
\end{subequations}
Thus, if  we were to describe our scatterer in terms of an ideal point-source monopole and dipole placed at the origin, we would find that the presence of the  aperture (i.e., the formation of a resonator) can be understood to cause these amplitudes to depend  upon both the incident pressure and the incident velocity  defined at the origin. Such cross-coupling is not observed in the analogous expressions for non-resonant cylinders: this is readily seen by considering the aperture closing limit $\varepsilon \rightarrow 0$  where $h_\varepsilon \rightarrow -\rmi \infty$ in \eqref{eq:allalphaslongform} which then recovers established results \cite[S59,S64]{melnikov2019acoustic}; in fact, the matrix in \eqref{eq:polarisabilitytensor} is strictly diagonal with $ {\alpha}_{12} = {\alpha}_{21} = 0$ and  $ {\alpha}_{22} = {\alpha}_{33}$ for    a closed Neumann cylinder.

\begin{figure}[t]
\centering
\subfloat[Subfigure 6 list of figures text][]{
\includegraphics[width=0.495\textwidth]{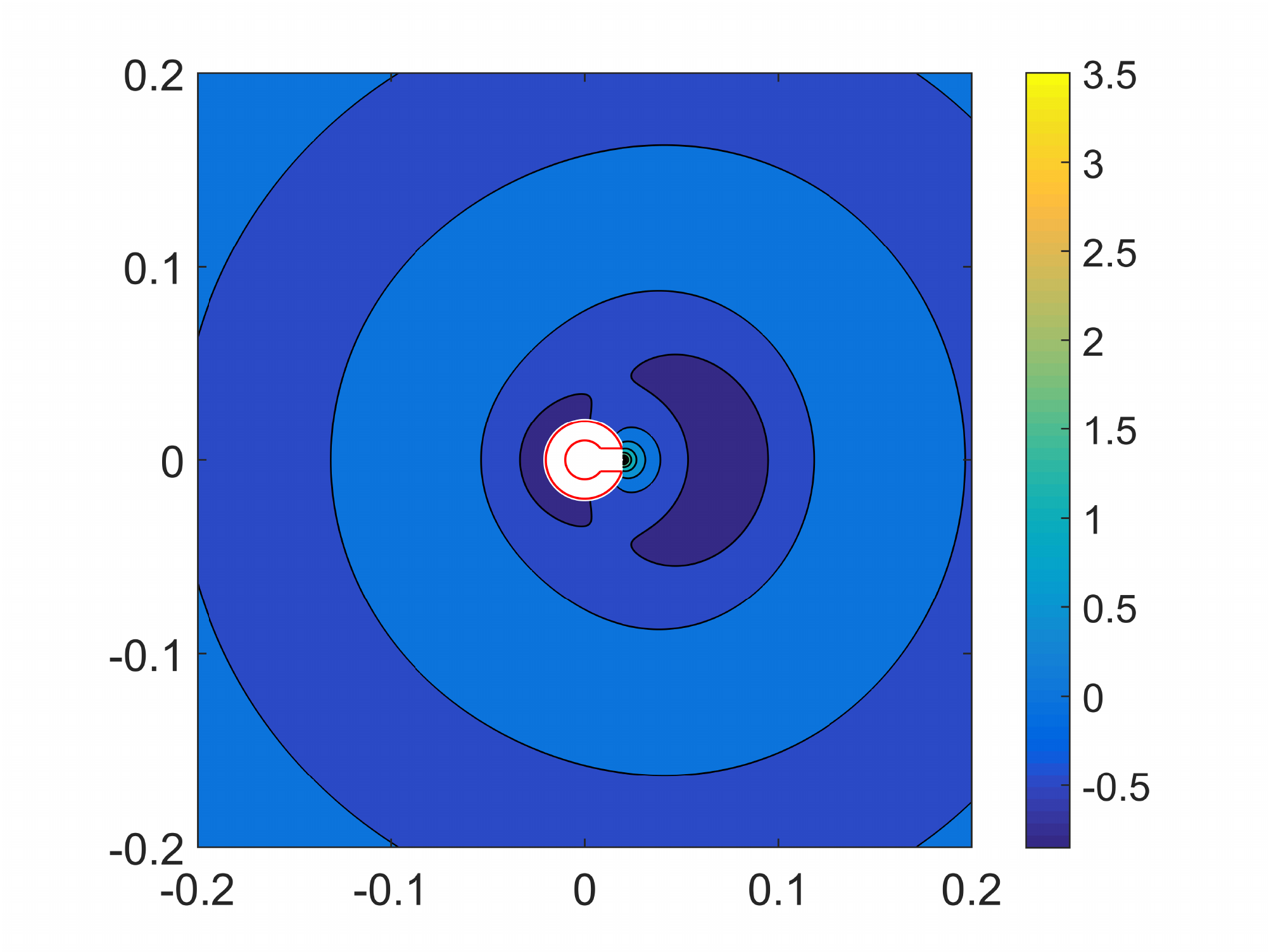}
\label{fig:figq1}}
\subfloat[Subfigure 1 list of figures text][]{
\includegraphics[width=0.495\textwidth]{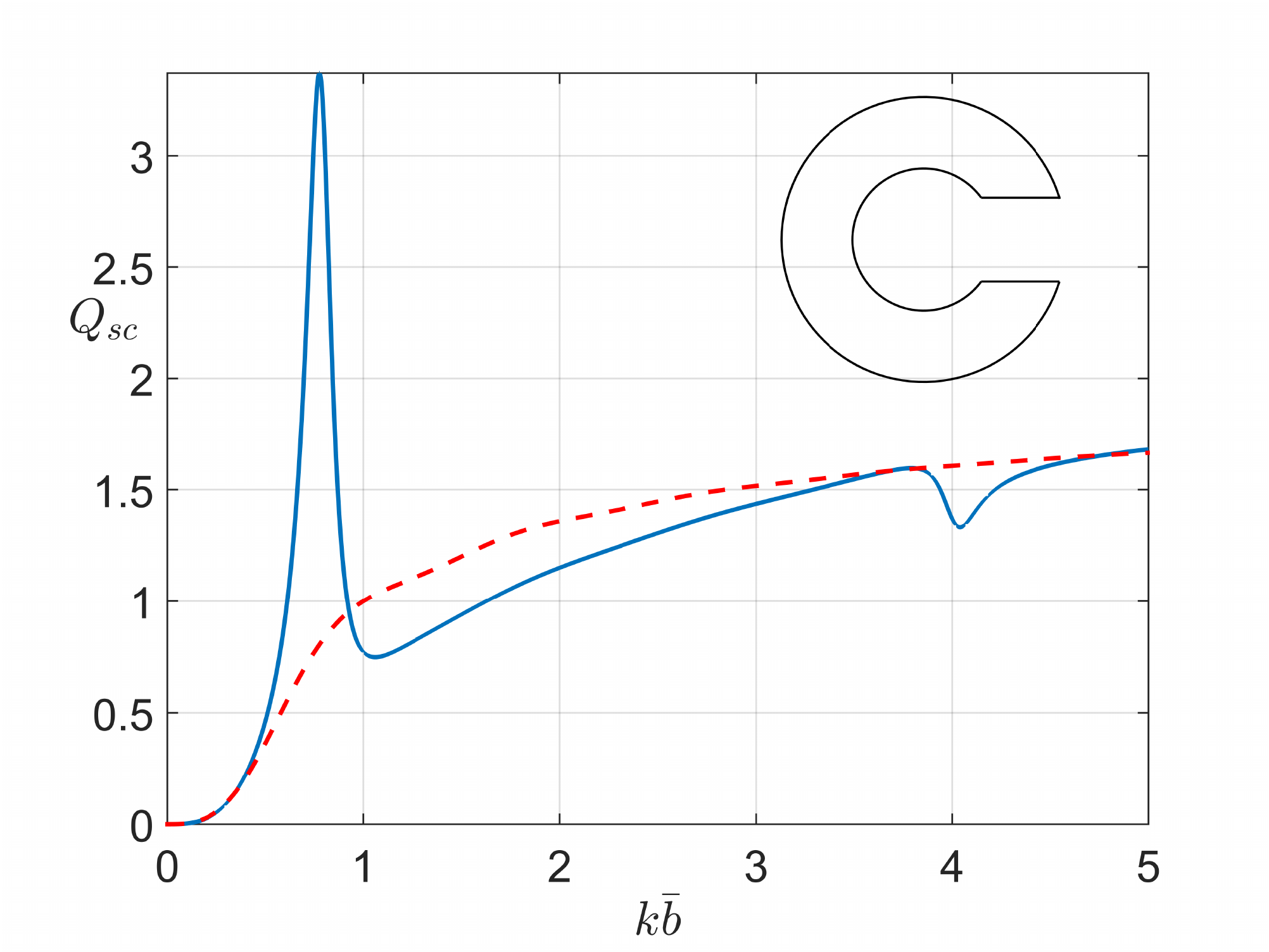}
\label{fig:figq2}}

\caption{Reference calculations for the configuration considered in Melnikov {\it et al.} \cite{melnikov2019acoustic}:   \protect\subref{fig:figq1}    the scattered potential $\phi_\mathrm{sc}$ as defined in \eqref{eq:ansatz1}  at $k=k_\rmH = 40.3589$ m$^{-1}$ and  \protect\subref{fig:figq2}    the (lossless)  scattering efficiency    $Q_\mathrm{sc}$ (defined from  \eqref{eq:sigmabarscatdef}).   In both figures we use   $\bar{b} = 20$\,mm, $2\bar{\ell} = 12$\,mm, $h=2\bar{\emm}/2\bar{\ell}=5/6$, and $\theta_\inc = \pi$; the result for a single closed Neumann cylinder of the same radius is superposed for reference (dashed red). Inset figure: resonator geometry (not to scale).}
\label{fig:alufigs}
\end{figure}

For the purposes of comparison, we present the scattered field profile $\phi_\mathrm{sc}$ and scattering efficiency $Q_\mathrm{sc}$ for the resonator geometry considered in Melnikov {\it et al.} \cite{melnikov2019acoustic} in Figure \ref{fig:alufigs}. Here, we observe  relatively strong   enhancements in both the scattered field and scattering efficiency at the Helmholtz resonance $k_\rmH$, as expected, and only comment that there could be room for  obtaining a stronger   response through an optimisation search of the type considered in Figure \ref{fig:figb4}.

Despite the emergence of off-diagonal terms   in the polarisability tensor,     earlier work by the authors and others \cite{smith2022PartII,schweizer2017resonance}, has found that the dispersion equation for a {\it two-dimensional square array} of cylindrical Helmholtz resonators possesses the usual structure for linear media, $\bar{k}_\rmB^2  \rho_\mathrm{eff}^{-1} - \omega^2 B_\mathrm{eff}^{-1} = 0$ at low frequencies, where the subscript eff denotes effective quantities and $\bar{\bfk}_\rmB$ denotes the dimensional Bloch vector. An example of a Willis coupling form  for an infinitely extending medium is given in \cite[Eq.~(2.68)]{quan2020nonlocality}, which along   an arbitrarily chosen high symmetry direction may take the form
\begin{equation}\label{eq:willisdispeq} 
\bar{k}_\rmB^2 - \omega (\zeta_\mathrm{eff} + \xi_\mathrm{eff}) \bar{k}_\rmB - \omega^2(\rho_\mathrm{eff} B_\mathrm{eff}^{-1}  - \xi_\mathrm{eff}\zeta_\mathrm{eff}) = 0,
\end{equation}
where $\zeta_\mathrm{eff}$ and $\xi_\mathrm{eff}$ are Willis coupling tensor elements. Upon introducing the conventional  decomposition $\zeta_\mathrm{eff} = \kappa_\mathrm{eff} + \rmi \chi_\mathrm{eff}$ and $\xi_\mathrm{eff} = \kappa_\mathrm{eff} - \rmi \chi_\mathrm{eff}$, where $\kappa_\mathrm{eff}$ and $\chi_\mathrm{eff}$ are the reciprocity and chirality parameters, respectively \cite{sieck2017origins}, the fact that the resonator geometry itself has at least one mirror plane of symmetry (i.e.,   is nonchiral) means that $\chi_\mathrm{eff} \equiv 0$. Similarly, if reciprocity were not preserved $\kappa_\mathrm{eff}\neq 0$, then a linear term in $\bar{k}_\rmB$ would be present in \eqref{eq:willisdispeq}. Since the effective dispersion equation   corresponding to a {\it two-dimensional array} of Helmholtz resonators does not possess  linear $\bar{k}_\rmB$ terms at low frequencies \cite{smith2022PartII}, and the resonator is nonchiral, we deduce that Willis coupling is therefore absent in bulk. Hence, it is unclear how the Willis-like coupling observed in the polarisability tensor \eqref{eq:polarisabilitytensor} for   a single resonator  implies the presence of Willis coupling effects in bulk, as is implied elsewhere, and would therefore encourage the use of alternative nomenclature (i.e., Willis-like depolarisability) for this behaviour. For reference, should $\kappa_\mathrm{eff} = 0$ then \eqref{eq:willisdispeq} takes the form
  \begin{equation}\label{eq:willisdispeq2} 
\bar{k}_\rmB^2   - \omega^2(\rho_\mathrm{eff} B_\mathrm{eff}^{-1}  - \chi_\mathrm{eff}^2) = 0,
\end{equation}
 from which it is  clear how   Willis coupling coefficients could be misattributed to the effective phase velocity; since our resonator geometry is nonchiral however we discount that possibility here.
  
\section*{Acknowledgements}
 I.D.A. gratefully acknowledges support from the Royal Society for an Industry Fellowship with Thales UK. This work was also supported by EPSRC grant no EP/R014604/1 whilst I.D.A. held the position of Director of the Isaac Newton Institute Cambridge. The authors acknowledge, with thanks, the invaluable assistance of  R. Assier  of the University of Manchester  in the development of the matched asymptotic expansion methodology for problems of this class.

\end{document}